# Development of Tool for Mapping Conventional Circuit to Reversible Logic

*A*

***Dissertation***

*Submitted in*

*Partial fulfillment*

*for the award of the Degree of*

***MASTER OF TECHNOLOGY***

***in Department of Computer Science Engineering***

***(With specialization in COMPUTER SCIENCE & ENGINEERING)***

**Supervisor**

Dr. S.C. Jain
(Professor)

**Submitted By**

Piyush Gautam
Enrolment No.:
10E2UCCSM4XT607

**DEPARTMENT OF COMPUTER SCIENCEENGINEERING**
**UNIVERSITY COLLEGE OF ENGINEERING**
RAJASTHAN TECHNICAL UNIVERSITY
KOTA (RAJASTHAN)

**March 2014**

# CANDIDATE'S DECLARATION

I hereby declare that the work, which is being presented in the Dissertation, entitled **"Development of Tool for Mapping Conventional Circuit to Reversible Logic"** in partial fulfillment for the award of Degree of **"Master of Technology"** in Dept. of Computer Science Engineering with Specialization in Computer Science, **and submitted to the Department of Computer Science Engineering,** University College of Engineering, Kota, Rajasthan Technical University is a record of my own investigations carried under the Guidance of **Dr. S.C. Jain**, Department of Computer Science Engineering, University College of Engineering, Kota**.**

I have not submitted the matter presented in this Dissertation anywhere for the award of any other Degree.

**Piyush Gautam**
Computer Science & Engineering
Enrolment No.: 10E2UCCSM4XT607
University College of Engineering,
Kota (Rajasthan)

Under Guidance of

**Dr. S. C. Jain**
Professor,
Department of Computer Science & Engineering
University College of Engineering,
Kota (Rajasthan)

# CERTIFICATE

This is to certify that this Dissertation entitled **"Development of Tool for Mapping Conventional Circuit to Reversible Logic"** has been successfully carried out by **Piyush Gautam** (Enrolment No.:10E2UCCSM4XT607), under my supervision and guidance, in partial fulfillment of the requirement for the award of **Master of Technology** Degree in **Computer Science & Engineering** from **University College of Engineering**, Rajasthan Technical University, Kota for the year 2010-2012.

**Dr. S. C. Jain**
Professor,
Department of Computer Science & Engineering
University College of Engineering,
Kota (Rajasthan)

## ACKNOWLEDGEMENTS

It is matter of great pleasure for me to submit this report on dissertation entitled **"Development of Tool for Mapping Conventional Circuit to Reversible Logic"**, as a part of curriculum for award of "Master in Technology" with specialization in "Computer Science & Engineering" degree of Rajasthan Technical University, Kota.

I am thankful to my dissertation guide **Dr. S.C. Jain**, Professor in department computer science for his constant encouragement, able guidance and for giving me a platform to build by career by giving me a chance to learn different fields of this technology. I am also thankful to **Mr. C.P. Gupta**, Associate Prof. & Head of Computer Science Department for this valuable support.

I would like to acknowledge my thanks to entire faculty and supporting staff of Computer Engineering Department in general and particularly for their help, directly or indirectly during my Dissertation work.

I express my deep sense of reverence to my parents and family members for their unconditional support, patience and encouragement.

**Date** **Piyush Gautam**

# CONTENTS

# List of Figures

# List of Tables

# List of Algorithms

# List of Abbreviations

| | |
|---|---|
| BDD | Binary Decision Diagram |
| ROBDD | Reduced Ordered Binary Decision Diagram |
| MAJ | Majority in Place |
| UMA | Un-Majority And Add |
| PPRM | Positive Polarity Reed Muller |
| NFT | New Fault Tolerant |
| IG | Islam Gate |
| PPTG | Positive Polarity Toffoli Gate |
| TIG | Two Inverter Gate |
| DFG | Double Feynman Gate |
| RPGA | Reversible Programmable Gate Array |
| ITRS | International Road-map for Semiconductors |
| IRC | Irreversible Circuit |
| RC | Reversible Circuit |
| BLIF | Berkeley Logic Interchange Format |
| ESOP | Exclusive-or Sum of Product |

# ABSTRACT

In the last decades, great achievements have been made in the development of computing machines. However, due to exponential growth of transistor density and in particular due to tremendously increasing power consumption, researchers expect that "Conventional Technologies" like CMOS will reach their limits in near future. To further satisfy the needs for more computational power, speed, less size etc. alternatives are needed. Reversible computation is the emerging field and alternative of conventional technologies.

Reversible computation is emerging as a promising solution and likely to work on extremely low power technologies and offer high speed computations. The reversibility retains the capability to retrieve the input data from output and minimizes heat dissipation.

As migration to new technology leave a lot of work done in current technology will make the acceptability difficult. One side familiarly with new technology and other side transformation of old circuit designs to new technology will pose a challenge to designers. A need for convertibility of irreversible circuit to reversible circuit was felt that can make a quick start and keep the development on track.

In this dissertation a logic circuit design energy based on binary logic system has been taken up that can provide the ease of circuit design in binary logic system and output as reversible circuit. Entire environment is GUI based and easy to learn to user friendly. The tool named "IRC2RC" offers editing, storage and conversion into reversible facility.

## Chapter-1

# INTRODUCTION

The evergrowing demand of high end computing applications have posed the challenge of continuous technology upgradation. The upgradation in technology has enabled the complex applications like Cloud computing, Real-time transitions on huge databases, Biotechnological computations a reality. Technological advancements in terms of higher operational frequency and miniaturization of chip in recent years have generated sufficient computing power to enable this growth. As predicted by Gordon Moore in 1960, popularly known as Moore's law, the transistor count in a chip will be double every one and half year on the average. Transistor growth is shown by Gordon Moore in figure 1.1. ITRS (International Technology Roadmap for Semiconductors) has also drawn a road-map of required feature size in future at atomic level in 2050 [1]. Shrinking in feature size resulted in a number of implementation and operational difficulties like heat dissipation, requirement of very thin laser beam, clock distribution etc.

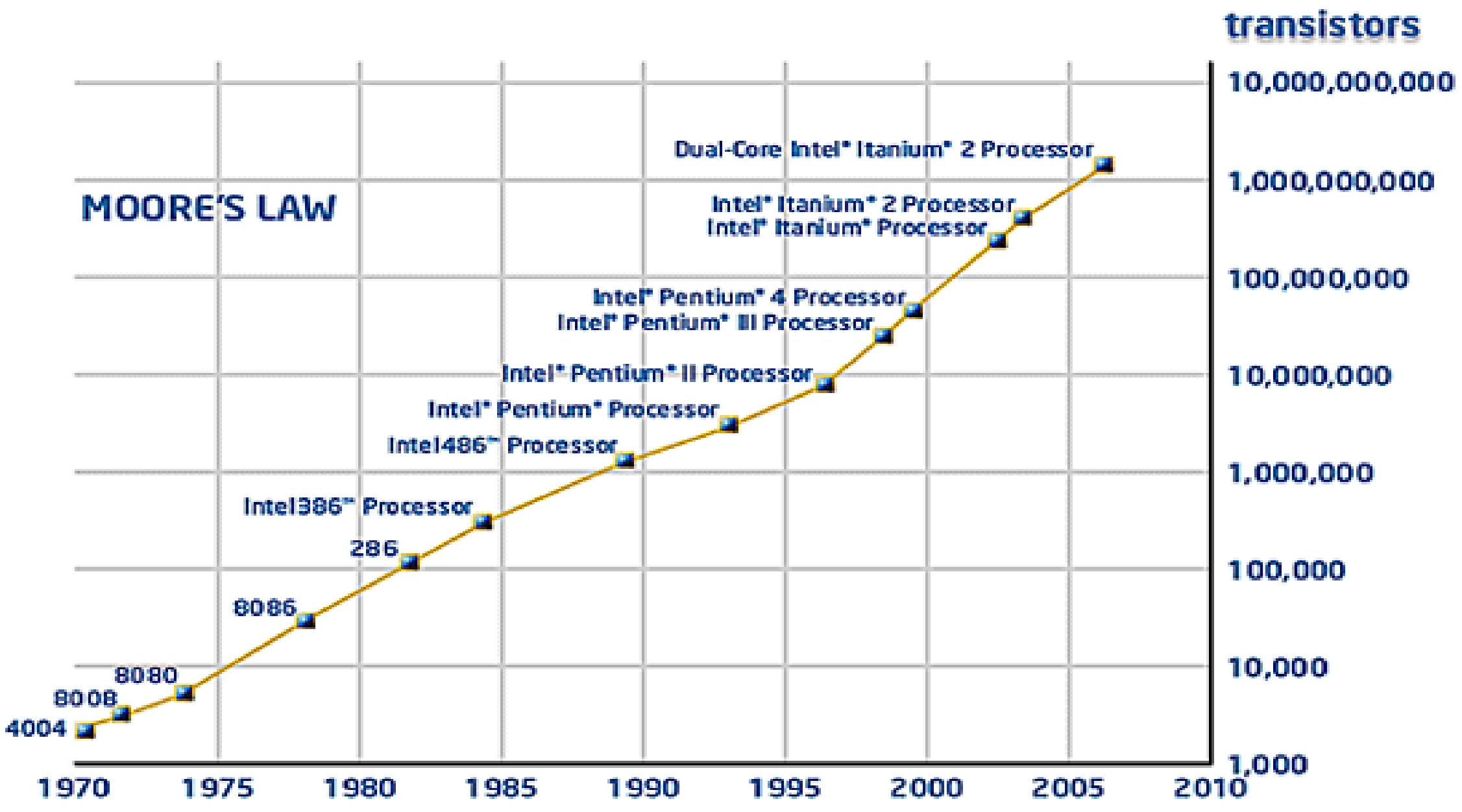

Fig 1.1 Moore Law's Graph for Transistors

Current technologies are finding it difficult to continue with the required level of growth. Alternative technologies are emerging to take place so that the growth momentum can be continued. Reversible computing is one of the computing system in which new generation computing system can be designed. Because of its basic nature of reversibility, it retains the old information and reduces dissipation of heat in its operation. This promise makes the technology as one of the possible alternative for future. This chapter describes the conventional computation and its limitations followed by reversible computing as an alternative.

## 1.1 Limitations of Conventional Computing

Conventional technology has dominated the computing world for more than last three decades. Right from the basic gates like AND, OR, NOT, EXOR, NAND etc. multimillion gates circuits have become available as per the need of the applications. The great deal of success has come from the fact that CAD tools, and VLSI technology for miniaturization have enabled the developments of large number of circuits. As a result complete process of conceptualizations till productions of chip has been well established and developing combinational and sequential circuits with fault tolerance capability have become easier.

The conventional computations by its nature are irreversible. The input cannot be reconstructed from its output. All the input lines do not propagate till output, resulting bit reversal or disappearance before output. This bit reversible causes a number of problems in high speed computations. That limits the viability of conventional computing for next generations.

Important resources which are involved in computing are Time, Space, Manufacturing cost and Energy. With growing demand of computational speed in scientific applications, it has been observed that irreversible behavior of classical gates may not be a technology rather it will lead to a number of problems [2]. The Problems, which may arise in conventional computing system, can be classified in following types.

1. Physical problems
2. Computational problems
3. Economic problems

### 1.1.1 Physical Problems

Irreversible classical gates based devices may become cause of physical inefficiently of conventional computing system.

#### 1.1.1.1 Heat Dissipation

According to Landauer (1961), using conventional (irreversible) logic, gate operation always leads to energy dissipation regardless of underlying technology. More precisely, kT.Ln2 Joule of energy is dissipated in each "lost" bit of information during irreversible computation, where k is Boltzmann constant and T is system Temperature. If T=300k which is equal to room temperature than heat dissipated is equal to $2.8*10^{-28}$ Joule/transistor. [3, 4]

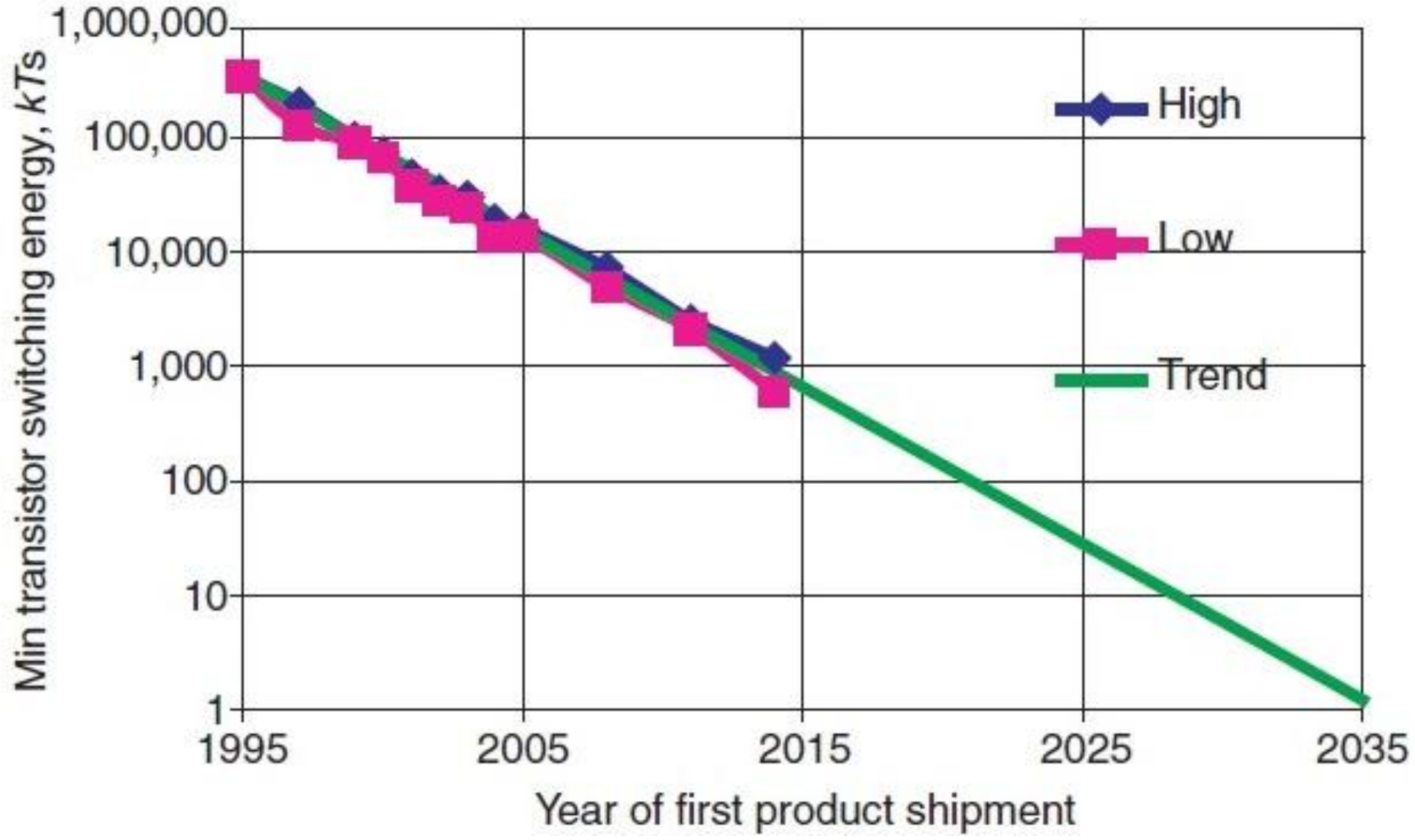

.

Fig 1.2 Trend of Minimum Transistor Switching Energy [3]

Energy is expressed as a multiple of room-temperature kT, which is also proportional to the number of nets of associated information. If the trend is followed, thermal noise will begin to become significant in the 2035s, when transistor energies approach small multiples of kTs. However, if reversible operations are used, the order-kT bit energies need not be dissipated, and so the dissipation per reversible bit manipulation might continue decreasing [4].

#### 1.1.1.2 Unable to Meet Size Requirements

According to Moore's law computational (Transistor) complexity is growing and feature size is shrinking according to ITRS (international Technology Roadmap for Semiconductor) feature size projection as per fig 1.3.

Nowadays computer are based on silicon chips. As chips become smaller and faster, chip packaging density increases. It is expected that required feature size will reach to atomic level in 2050, but due to heat dissipation we cannot achieve high packaging density as we have limit on distance between adjacent bit devices, so conventional computers will not be able to meet size requirements in coming years.

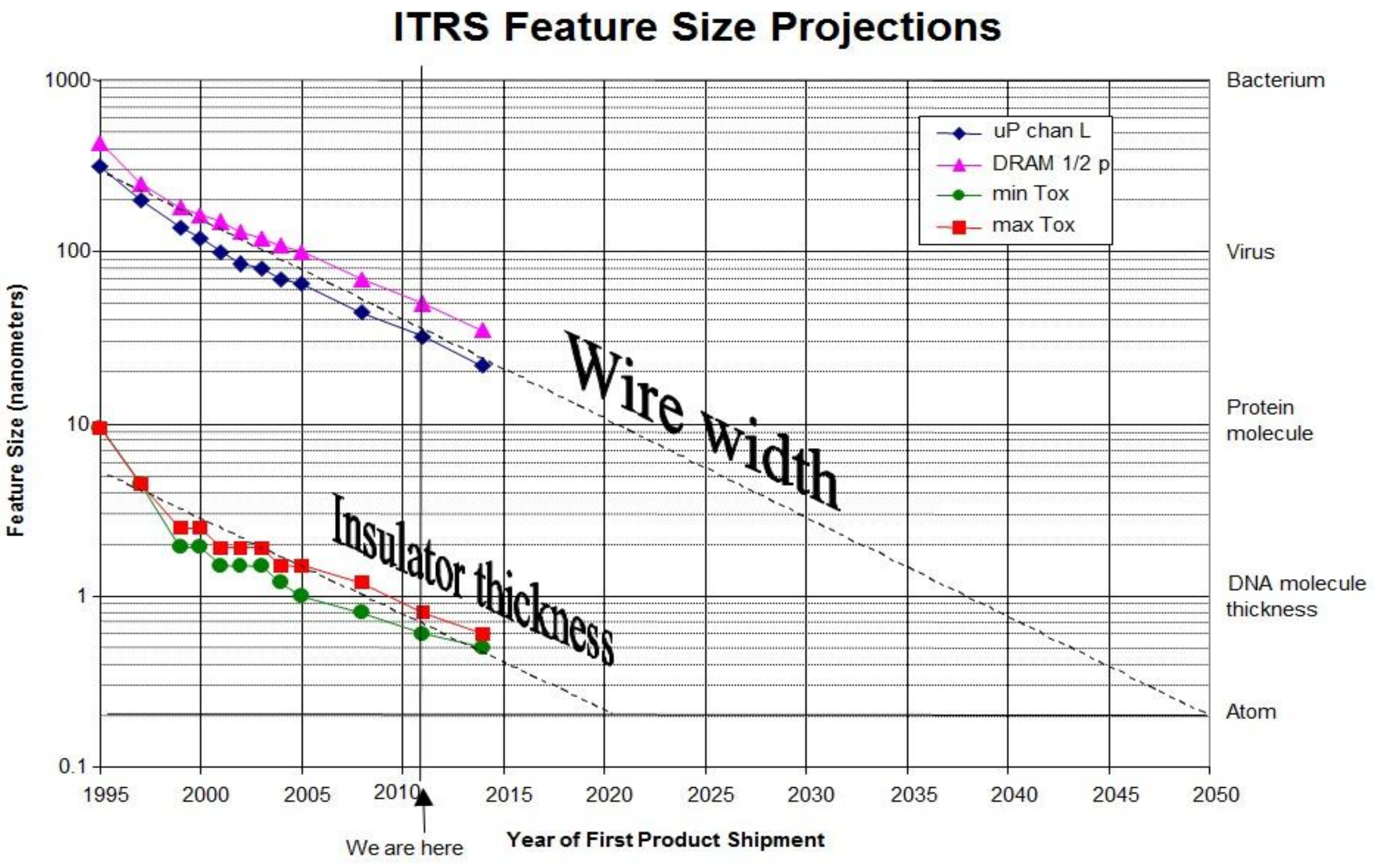

Figure 1.3: ITRS Feature Size Projection

#### 1.1.1.3 Operational Speed

Conventional computers are facing problems to meet the demand of higher operational speed for applications like Cloud computing, Real time applications and other scientific applications. Due to limit on speed of light, heat dissipation, stray capacitances etc. Classical computer systems are not able to achieve the required computational speed and clock frequency.

### 1.1.2 Computational Problems

A large number of computational intensive problems like NP-Complete type problems demand high computational speed, but have not been solved by classical computers. In particular backtracking problems in complex applications are highly composite intensive require high speed. Also cryptanalysis methods that are based on heat dissipation during cryptography operations pose a serious security threat.

### 1.1.3 Economical Problems

Apart from technical problems, cost of computation will not be reduced with higher complexity. Hence higher computational complexity will no longer remain economical.

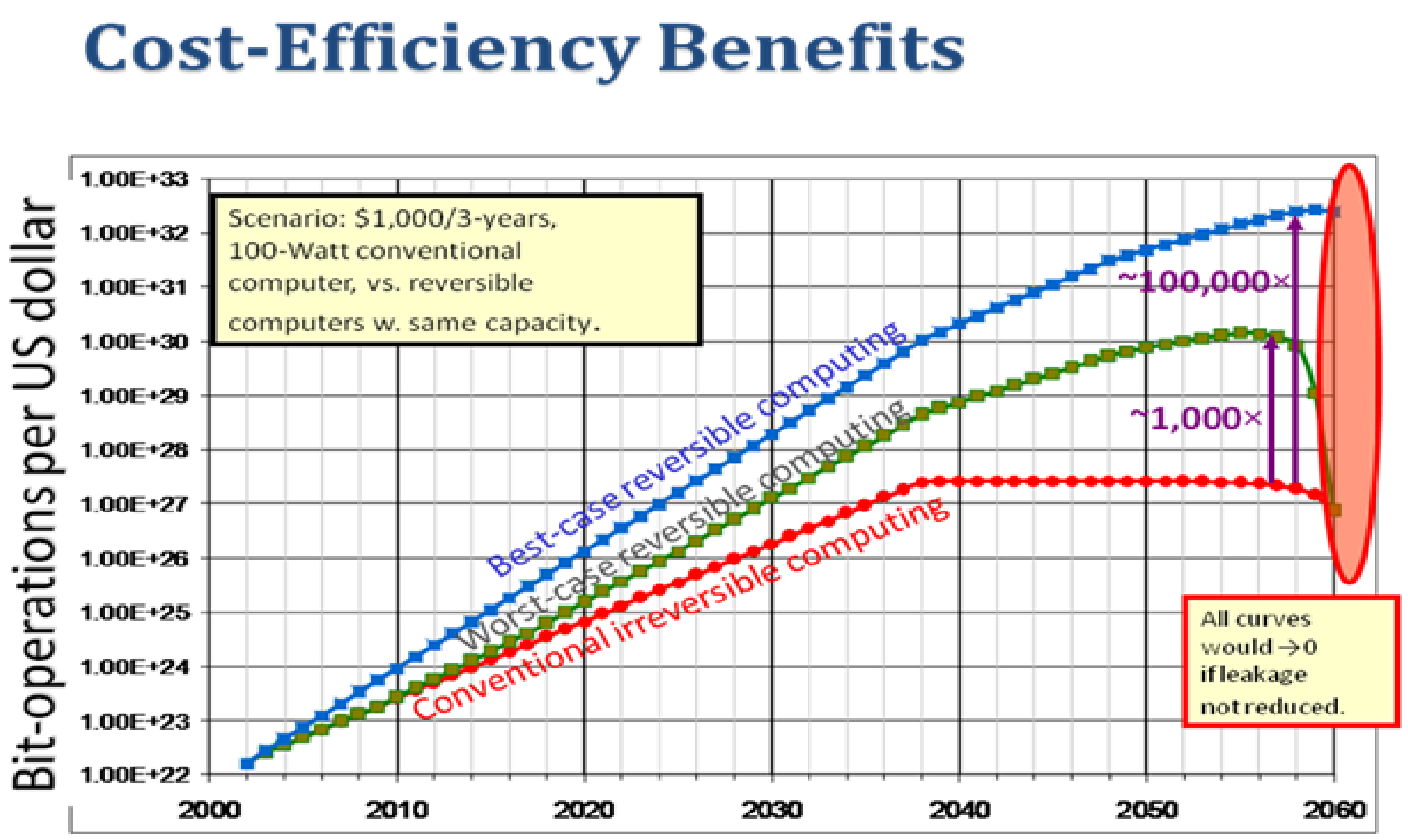

Fig 1.4 Cost Efficiency Benefits [5]

Figure 1.4 shows the number of bit operations per dollar of irreversible computations. It is likley to remains stagnant, where as reversible computing is likely to offer significant advantage.

## 1.2 Reversible Computation

Due to the limitations of conventional computing cited above, reversible computation offers a possible solution. The reversible computation saves heat dissipation by avoiding bit destruction and hence preserves reversibility.

In mid-60 and 70 researchers consider this problem as, whether it is possible to compute without generating heat. Landauer's principle states that the source of heat generation computation is destruction of bits of information, not their transformation [2]. Bennett [1961] showed that energy dissipation can be reduced or even eliminated if computation becomes loss–less [4]. This does not hold for conventional circuits (because they lead to loss of information) but in reversible circuits where all the operations are performed in invertible manner, and satisfy this criteria of loss–less computation.

The above problem can be eliminated form system by using Reversible Logic, which can be performed by using reversible gates and is known as reversible computing. The reversible computation is simply based on the fact that, the existing information in any system can’t just be destroyed. This is because at the lowest level physics is reversible that means in a closed system it transforms one state to another over time in mathematical invertible way.

To achieve reversibility we have to design logics such that retrieved output form the logics should have sufficient information to get back inputs. In computational system we should have gates which have one to one mapping between input and output, that computational system is called reversible computation. Reversible logic does not result into destruction of information (bits) but it just map one state to other. To achieve this, reversible logics are designed with same number of input and output. Reversible gates are gates which uses reversible logics and become there is no loss of information hence reversible gates does not results into any heat dissipation.

Moreover, as shown by Bennett, zero power dissipation on circuit will only be possible if the respective computation is reversible. Because of less or no heat dissipation we can achieve high density and so we can achieve smaller size. Hence, reversible circuits are seen as future alternative to conventional circuit technology with certain low power applications.

Performing computation with reversible logics is called reversible computing, reversible computing would require new software and hardware design, development tool, work with algorithms that don’t require data to be erased, programming language and compilers etc. But, in order to reuse the circuits already developed in irreversible computing, re-development of circuits is required. The advantage of established tools may also not be available.

## 1.3 Objectives and Motivations

As we feel that the field of reversible computing is emerging and promising in which few tools are available and future computers are likely to be based on such technology. Large number of circuit has been developed in conventional designs and a number of synthesis tools are available to develop conventional circuits. Switching to new technology will require fresh design of the same circuits which will be a tedious task. A conversion form conventional to reversible circuit will not be that easy. Finding adequate number of input and outputs and mapping to bijective functions will require appropriate technology mapping. We undertake the project with this objective. We undertake the project develop a tool that maps a conventional circuit to its equivalent reversible circuit.

## 1.4 Organization of Dissertation

**Chapter 2:** The chapter with the name “LITERATURE SURVEY” will describes literature survey about reversible computation which will help the readers to understand the basic and progress of different areas in the field of reversible logic.

**Chapter 3:** The chapter with the name “CONVERSION APPROACH AND DESIGN ALGORITHMS” of this dissertation report is going to explain the conversion approaches and methodology with algorithms used to implement this work.

**Chapter 4:** The chapter with the name “INTEGRATION AND TESTING” is made for the purpose of graphical representations and testing results.

**Chapter 5:** Finally this dissertation work is concluded with specific contributions and giving ideas of future extension of this dissertation work in the chapter named “CONCLUSION AND DIRECTION OF FUTURE WORK”.

# Chapter 2

# LITERATURE SURVEY

A literature survey is a critical and in depth evaluation of previous research in the area of reversible computing. Reversible computing is emerging as a potential development platform to replace conventional logic. This chapter represents previous work on reversible logic. We categorize our survey in the following categories.

- Reversible logic gates
- Circuit formats
- Reversible circuit design
- Tools
- Others

At the end of survey we will analyze potential problem for our work.

## 2.1 Reversible Logic Gates

Right from the stored program architecture given by John Von Neumann in 1949, heat dissipation per computation of bit is being estimated. R. Landauer [1961] pointed out that the irreversible erasure of a bit of information consumes power and dissipates heat. While reversible designs avoid this aspect of power dissipation. Destruction of bits causes heat dissipation as per Landauer Principle[3]. Bannett in 1973 proposed a turing machine for loss-less compuation by making it reversible[7]. The development of reversible gates and circuits started after Toffoli proposed reversible logic gates in 1977[8]. A number of gates have been proposed thereafter. The same has been described two categories namely basic gates and generalized gates.

### 2.1.1 Basic Gates :

A reversible gates realize a reversible function, computation done by a gate is reversible in nature, that means for a gate g the gate $g^{-1}$ implement inverse transformation . Some of the basic gates are given below:

A **Not gate** is a 1×1 reversible gate is shown in Fig. 2.1. The input is A and the output is P = A' which is reversible.

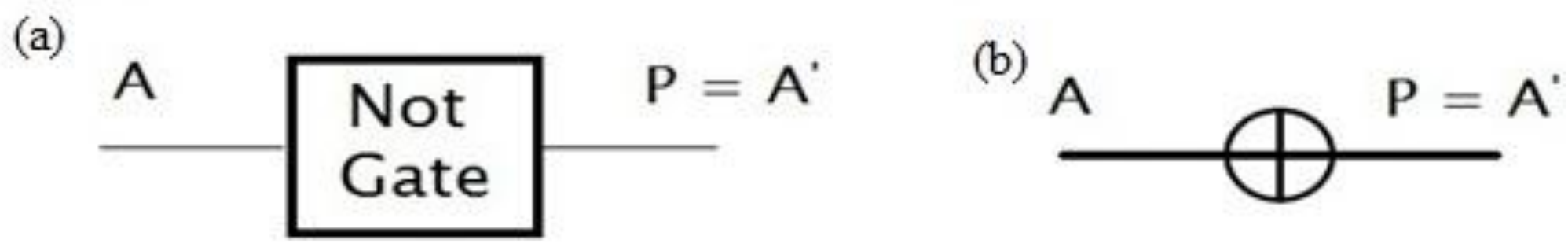

Fig 2.1 Reversible NOT Gate (a) Block Diagram (b) Schematic Representation

Table 2.1 Trurh Table of NOT Gate

| Input | Output |
|---|---|
| A | P |
| 0 | 1 |
| 1 | 0 |

**Feynman gate** was given by Richard Feynman in 1982, which is a 2×2 reversible gate. Feynman gate can perform negation operation but in controlled way and it is also known as Controlled NOT Gate. If two line are A and B, the first line A is known as CONTROL line and second line B is known as TARGET line. Operation on target line is negation and only performed when control line is set otherwise no operation on target line is observed.

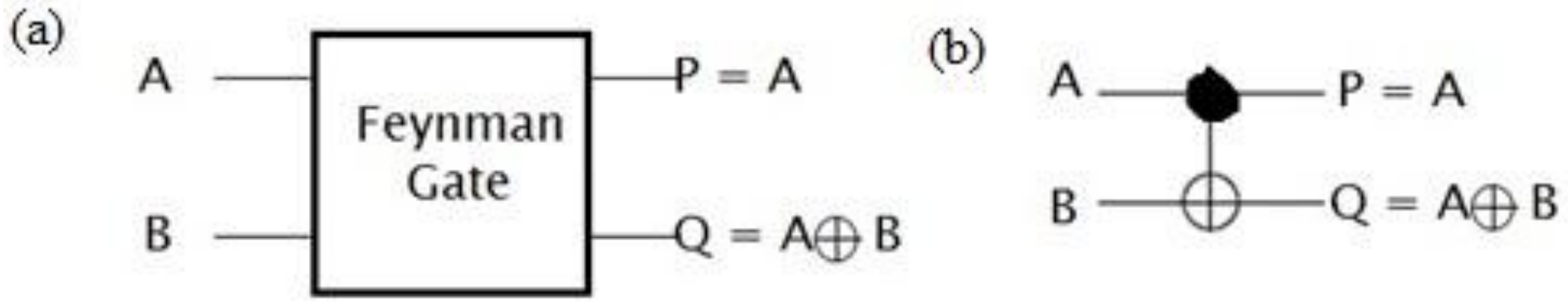

Figure 2.2 Feynman Gate (a) Block Diagram (b) Schematic Representation

Table 2.2 Truth Table of Feynman Gate

| Input | | Output | |
|---|---|---|---|
| A | B | P | Q |
| 0 | 0 | 0 | 0 |
| 0 | 1 | 0 | 1 |
| 1 | 0 | 1 | 1 |
| 1 | 1 | 1 | 0 |

In 1982 Toffoli give a new gate called **Toffoli gate**. It is a 3×3 gate and can be generalized up to n×n size[9]. As per definition target line flips when all control lines are set. Fig 2.3 shows  block diagram and schematic representation of 3×3 toffoli gate.

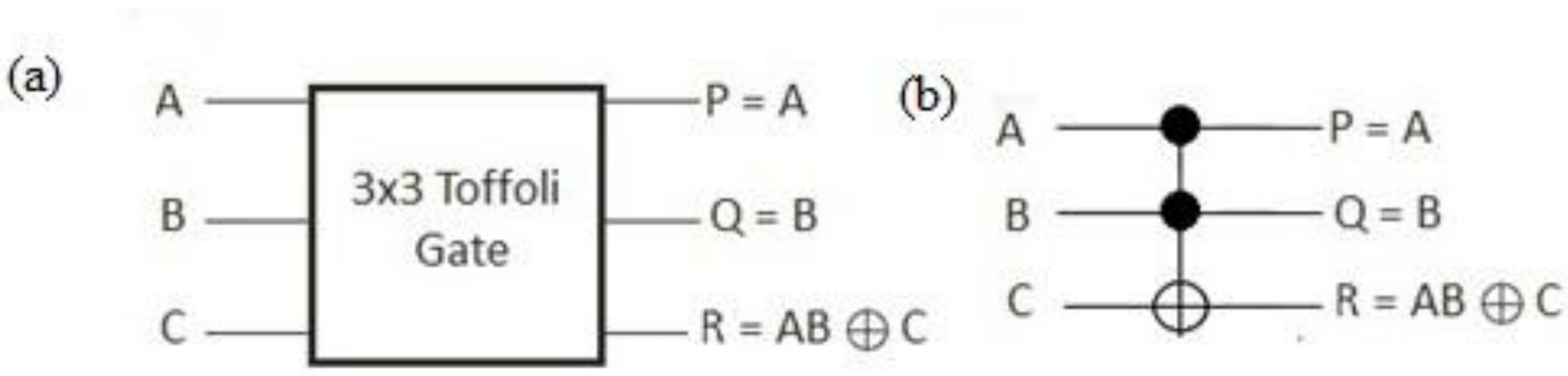

Fig 2.3 3×3  Toffoli Gate (a) Block Diagram (b) Schematic Representation

Table 2.3 Truth Table of Toffoli Gate

| Input | | | Output | | |
|---|---|---|---|---|---|
| A | B | C | P | Q | R |
| 0 | 0 | 0 | 0 | 0 | 0 |
| 0 | 0 | 1 | 0 | 0 | 1 |
| 0 | 1 | 0 | 0 | 1 | 0 |
| 0 | 1 | 1 | 0 | 1 | 1 |
| 1 | 0 | 0 | 1 | 0 | 0 |
| 1 | 0 | 1 | 1 | 0 | 1 |
| 1 | 1 | 0 | 1 | 1 | 1 |
| 1 | 1 | 1 | 1 | 1 | 0 |

In 1982 Edward Fredkin and Tommaso Toffoli proposed a new gate called 3×3 **Fredkin Gate**[10] which is further generalized up to n lines. Figure 2.1 shows the block diagram and schematic representation of 3×3 Fredkin gate. If C=0 3×3 Fredkin gate will swaps the values of A and B.

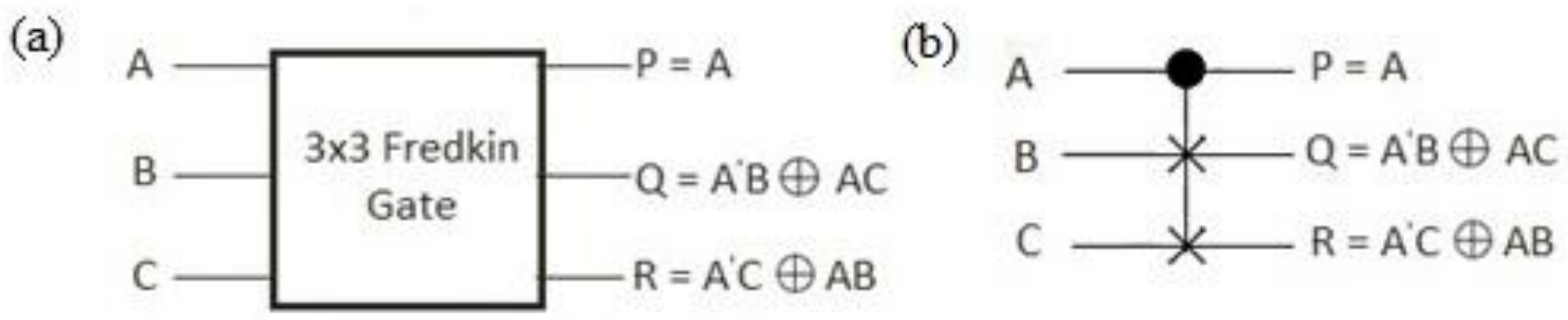

Fig 2.4 3×3 Fredkin Gate (a) Block Diagram (b) Schematic Representation

Table 2.4 Truth Table of Fredkin Gate

| Input | | | Output | | |
|---|---|---|---|---|---|
| A | B | C | P | Q | R |
| 0 | 0 | 0 | 0 | 0 | 0 |
| 0 | 0 | 1 | 0 | 0 | 1 |
| 0 | 1 | 0 | 0 | 1 | 0 |
| 0 | 1 | 1 | 0 | 1 | 1 |
| 1 | 0 | 0 | 1 | 0 | 0 |
| 1 | 0 | 1 | 1 | 1 | 0 |
| 1 | 1 | 0 | 1 | 0 | 1 |
| 1 | 1 | 1 | 1 | 1 | 1 |

Swapping is important phenomenon for logic gates and swapping is important for many systems. **Swap gate** is reversible gate which is basically 2*2 gate $S(x_1,x_2)$ which swap values of $x_1$ and $x_2$[10]. Figure 2.5 shows schematic representation of swap gate.

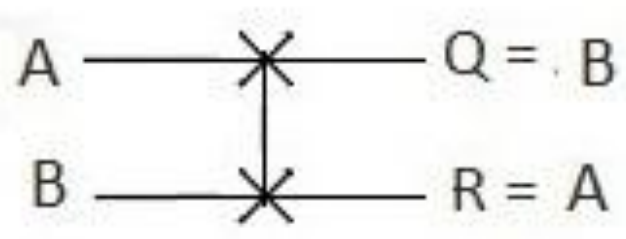

Figure 2.5 Schematic Representation of Swap Gate

Table 2.5 Truth Table of Swap Gate

| Input | | Output | |
|---|---|---|---|
| A | B | P | Q |
| 0 | 0 | 0 | 0 |
| 0 | 1 | 1 | 0 |
| 1 | 0 | 0 | 1 |
| 1 | 1 | 1 | 1 |

### 2.1.2 Generalized and Complex Gates :

In addition to basic gates, some generalized and complex gates have also been proposed in the literature.

- **Multi-Control Toffoli Gate:** In 1980 Toffoli given a gate that can be generalized up to n lines this can implement functionality of several gates[9]. Multi bit toffoli gates passes the first m lines unchaged ,which are called control line and flip the target line $((m+1)^{th})$ line if and only if each positive (negative) control line carries the 1(0) value for m=0,1,2 the gates are named NOT(N), CNOT(C) and Toffoli(T) respectively. These three gates compose the universal NCT library.

- **Multiple-Control Fredkin Gate :** In 1982 Toffoli and Fredkin given a gate that can be generalized up to n lines . Fredkin $(x_1,x_2,\ldots\ldots x_{m+2})$ has two target line $x_{m+1},x_{m+2}$ and m control line $(x_1,x_2,\ldots\ldots x_m)$.The gate interchanges the values and of the targets and if the conjunction of all m positive(negative) controls evaluates to 1(0). For m=0, 1 the gates are called SWAP(S) and Fredkin (F) gate respectively [9].

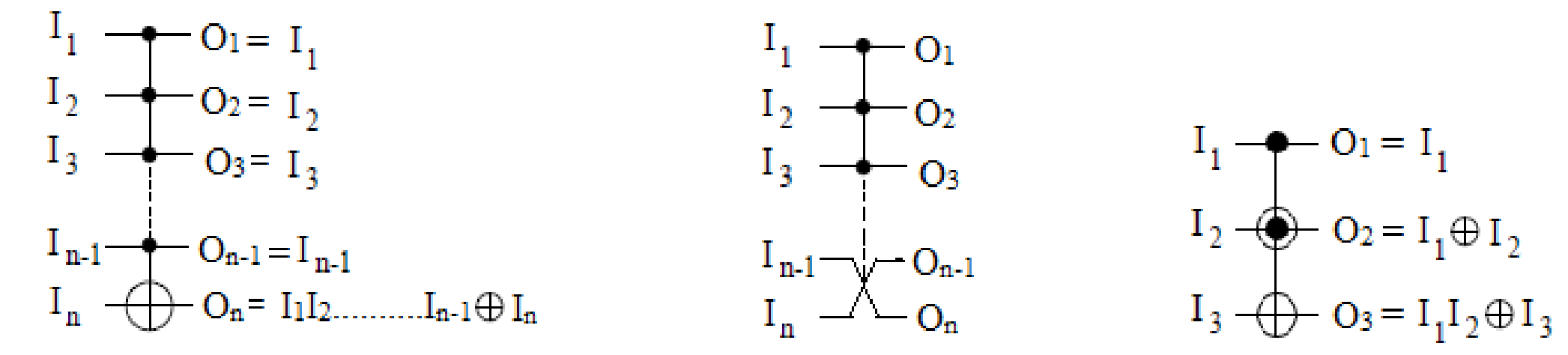

(a) n-bit MCT Gate (b) n-bit MCF Gate (c) Peres Gate

Figure 2.6 Generalized Reversible Gate

- **A Peres Gate:** Peres gate is proposed by A.Peres in 1985[11]. It is also 3×3 reversible gate and Peres gate P(x1,x2,x3) has one control line x1 and two target line x2 and x3. It represents a Toffoli(x1,x2,x3) and a Feynmen(x1,x2) in a cascade.

If the sequence is changed, the gate is called Inverse Peres Gate. Two cascaded Peres-gate works as half adder and if we put C=0, then it can realize AND operation at target line R[12].

- **MAJ Gate and UMA Gate:** In 2005 Cuccaro presents a Majority in place [MAJ] gate and Un-Majority and add (UMA). A majority in place [MAJ] gate computes the majority of three bits in place and provides the carry bit for addition. MAJ and UMA gates are basically made-up of two CNOT gate and one TOFFOLI gate. For MAJ and UMA gates, the inverse gates can be constructed by reordering the CNOT and Toffoli gates [13].

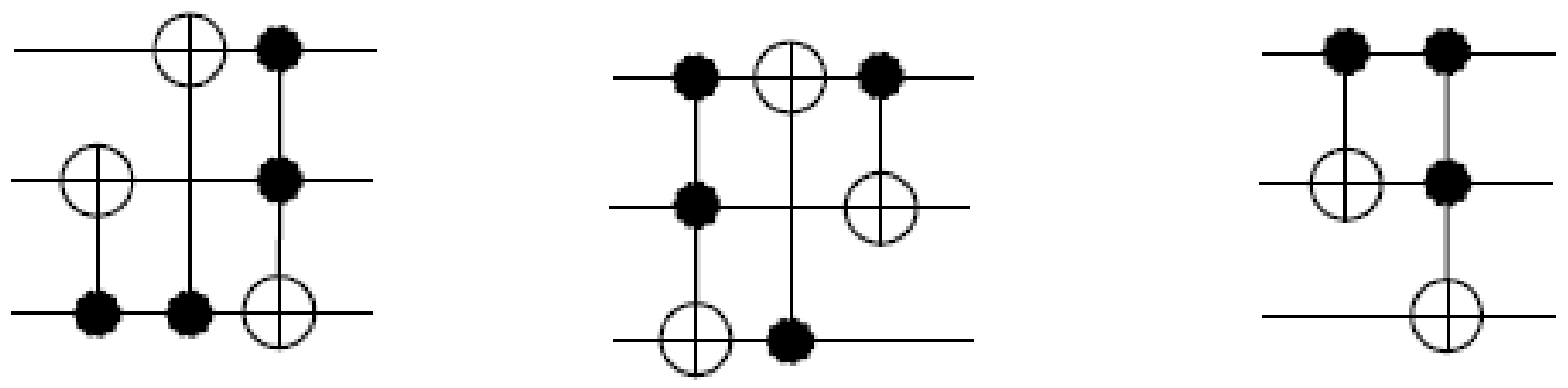

(a) MAJ Gate (b) UMA Gate (c) Inverse Peres Gate

Figure 2.7 Other Generalized Gates

The above gates are generalized and are used for desiging for specific applciation by the authors.

### 2.1.3 Fault Tolerant Gates:

Fault tolerance in reversible circuit is being proposed by making the gates parity preserving. In these gates, the parity of inputs makes with output. Parhami has given the concept of Parity Preserving Reversible Logics in 2006[14] that are reversible as well as parity.

➢ **F2G Gate:** In 2006 Behrooz Parhami presents a Feynman double gate in 2006, it is basically designed by 2 Feynman gates and so called Feynman double gate or F2G [15]. Quantum cost of double Feynman gate is 2. Figure 2.8 shows the Double Feynman gate.

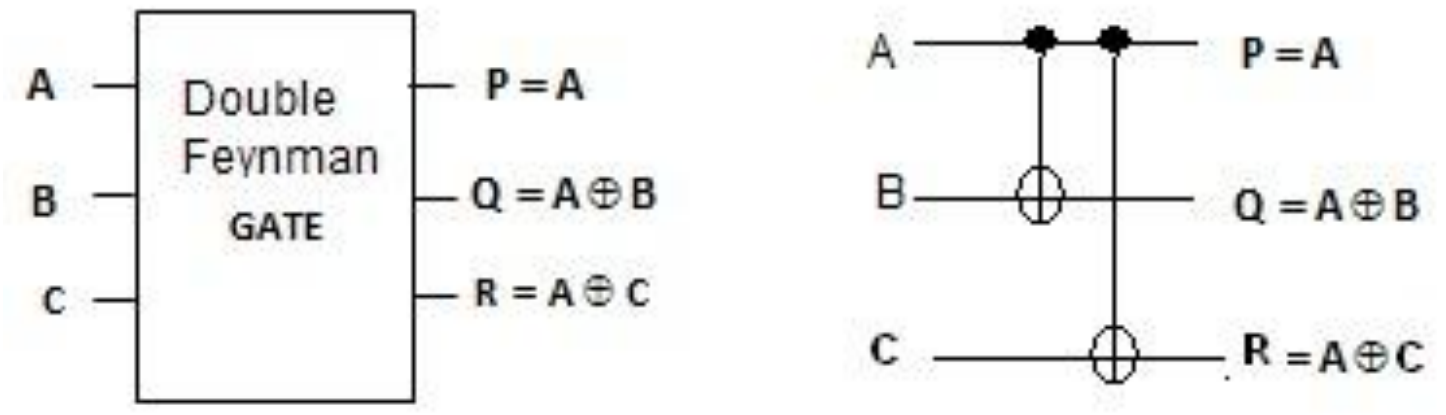

Figure 2.8 Double Feynman Gate

➢ **NFT Gate:** In 2008 Haghparast M presents a new gate called NEW FAULT TOLERENT Gate called NFT for nanotechnology based systems which satisfies the condition of parity preserving hence can be used a fault tolerant gate[16]. Figure 2.9 shows NFT gate.

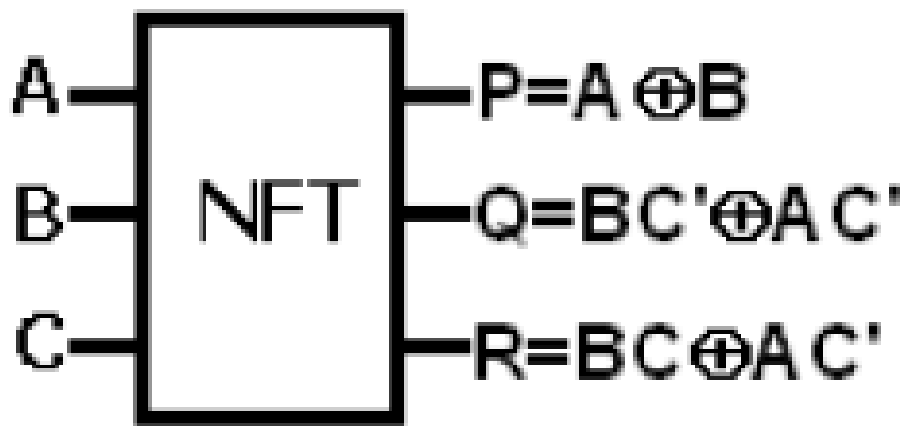

Figure 2.9 New Fault Tolerent Gate

➢ **IG Gate:** In 2010 Saiful Islam, Rehman, Hafiz and Zerina Begum have given design of a new fault tolerent reversible Full–Adder in which they proposed a new 4*4 gate with the name of Islam Gate (IG Gate). It is also a one– through gate [17], they have also shown that proposed IG Gate is universal gate because it can be used for implementing any arbitrary boolean function as shown in figure 2.10. Figure 2.10 shows the some conversion with IG Gate.

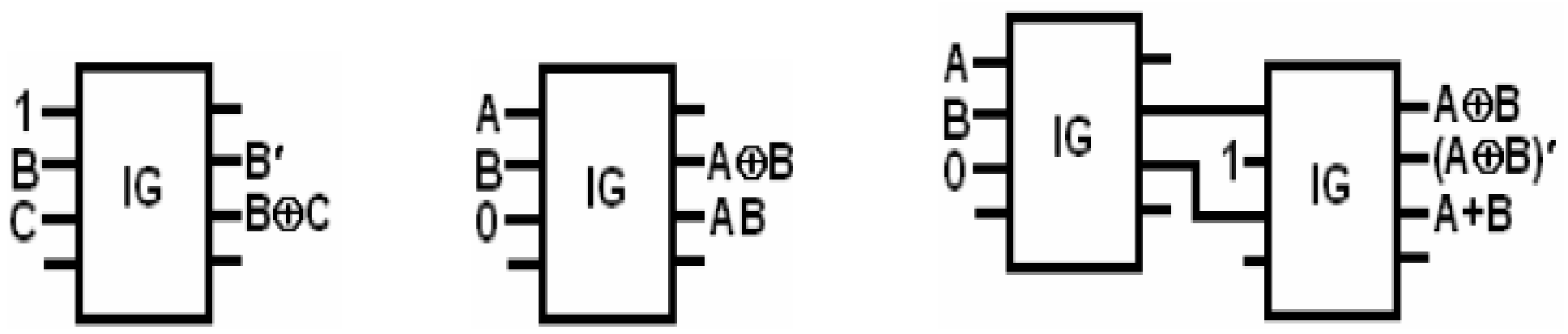

a. IG as Inverter and EX-OR. b. IG as AND and EX-OR. c. IG as EXOR, EX-NOR and OR

Figure 2.10 IG Gate as Universal Gate

- **PPTG(Parity-Preserving Toffoli Gate):** In 2013 Dr. S.C. Jain, Angurah Jain and Nitin Purohit proposed a 4*4 parity preserving toffoli gate. This gate is a Toffoli gate with one additional input and output. The additional input and output with existing control inputs A and B makes this gate versatile and reduces cost [18].Figure 2.11 shows the diagarm of PPT gate.

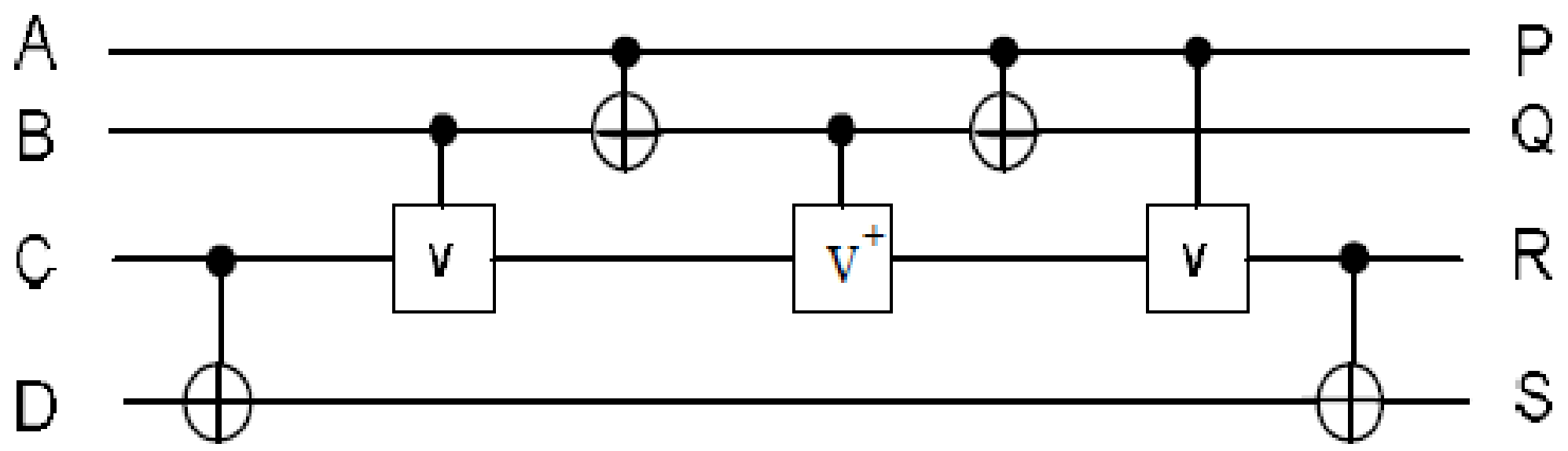

Figure 2.11 Parity-Preserving Toffoli Gate

- **TIG Gate:** In 2013 Dr. S.C. Jain, Angurah Jain and Nitin Purohit proposed a 2*2 parity preserving TIG(Two Inverter Gate). It inverts two input bits in the outputs. Therefore we are referring it as TIG (Two Inverter gate) [18]. TIG gate in block diagram and symbolic form are shown in Figure 2.12.

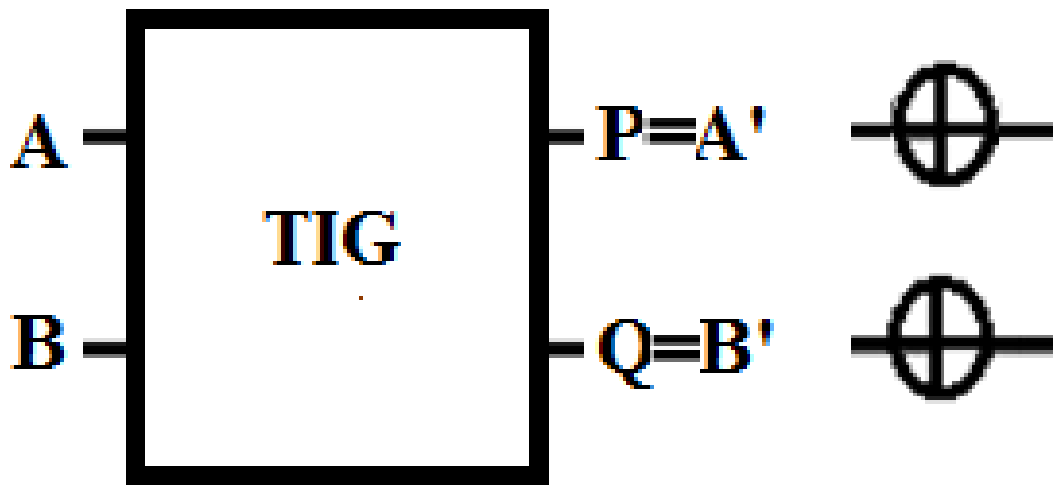

Figure 2.12 Parity Preserving TIG Gate

### 2.1.4 Reversible gates for Reversible Sequential Circuits

This section will describe some of the reversible gates presents specifically for the purpose of making and improving Sequential Reversible Circuits.

- **SG Gate:** In 2010 Abu Sadat and Masashi Ueda have proposed a new gate 4*4 called SG Gate for the purpose of optimizing design of Sequential elements (D-Latch and JK-Latch), newly proposed SG Gate is two universal gate [19].

- **SVS Gate:** In 2013 Dr. S.C. Jain, Shubam Gupta and Vishal Pareek have proposed a new gate 4*4 called SVS gate. This gate given a signifcant improvement in realizing and optimizing T flip-flop[20].

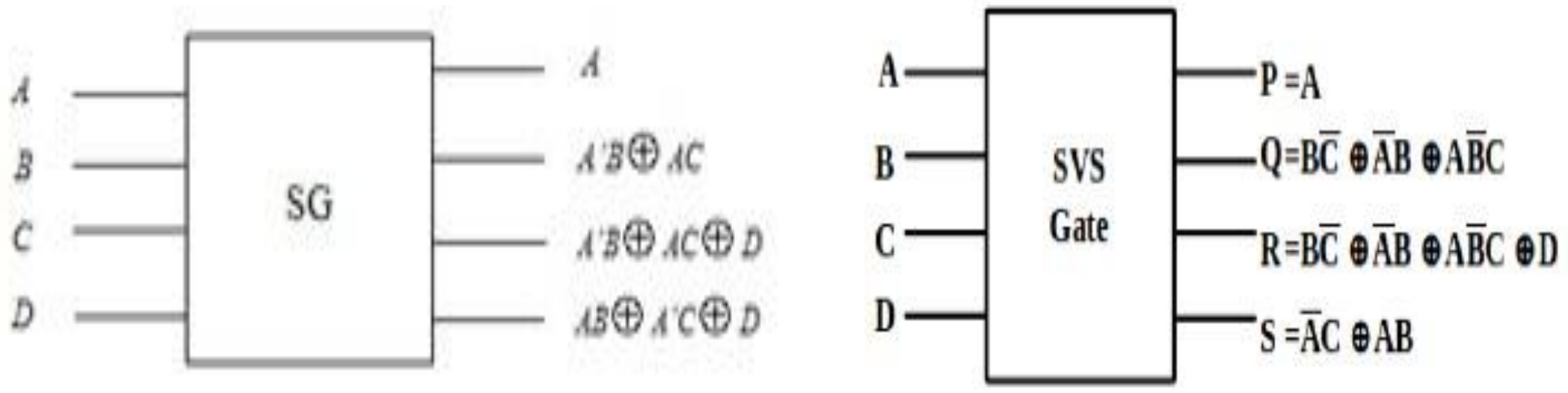

(a) SG Gate (b) SVS Gate

Figure 2.13 Reversible Gates for Reversible Sequential Circuits

## 2.2 Circuit Represenation Format

Reversible Circuit can be represented in the several ways and each format of representation can be used in different syntesis approaches. New methods have been proposed form 1986 to 2013 for representation of reversible functions. Some format of circuit representation here as follows:

**Truth Table** is one of the easiest method to describe a reversible function. Reversible truth table contains input vector including both primary and constant input and output vector including Garbage and Primary outputs. An irreversible truth table contains only primary inputs and corresponding primary outputs in each row. Reversible truth table was proposed by **Toffoli** and **Fredkin** in **1982**[10].

**Binary Decision Diagram** for a Reversible function can be represented by a Binary Decision Diagram(BDD). BDD is directed acyclic graph, Reduced Oredered Binary Decision Diagrams(ROBDDs) is a BDD, which offer canonical representations of Boolean Functions. ROBDD can be constructed from a BDD by ordering variables, merging equivalent sub-graphs and removing nodes with identical children. It was proposed by **R.E. Brayant** in **1986**[21].

**Cyclic graph** is one of the shortest format for representation of reversible circuits, it represents the cyclic chain of input and outputs and useful in cycle based synthesis approach for reversible circuits. The individual cycles are treated as building blocks of reversible circuits. It was proposed by **J.D.Dixon** and **B.Mortimer** in **1996**[22].

**Positive Polarity Reed-Muller Expansion(PPRM**) are used by Search based synthesis approach. In this approach any Boolean function can be represented using boolean variable and XOR operators. The Positive Polarity Reed-Muller Expansion(PPRM) uses only uncomplemented variables and can be derived easily from functions sum of product expansion. It was proposed by **T. Sasao** and **M. Fujita** in **1996**[23].

In **Matrix Represnetation** boolean reversible function f can be described by a Sparse matrix of 0 and 1 with single 1 in each column and in each row, where the non-zero element in row i appers in column f(i). This is also easy method to represent a reversible function, by using Matrix Representation one can easily analyze that reversible function have one to one mapping between input and output. It was proposed by **K.N.Patel** in **2010**[24].

| xyz | $f_1f_2f_3$ |
|---|---|
| 000 | 001 |
| 001 | 000 |
| 010 | 011 |
| 011 | 010 |
| 100 | 110 |
| 101 | 100 |
| 110 | 101 |
| 111 | 111 |

(a) Reversible Truth Table

(0,1)
(2,3)
(4,6,5)

(b) Cycle Form

$f_1 = x$
$f_2 = x \oplus y$
$f_3 = 1 \oplus x \oplus z \oplus xy \oplus xz$

(c) PPRM

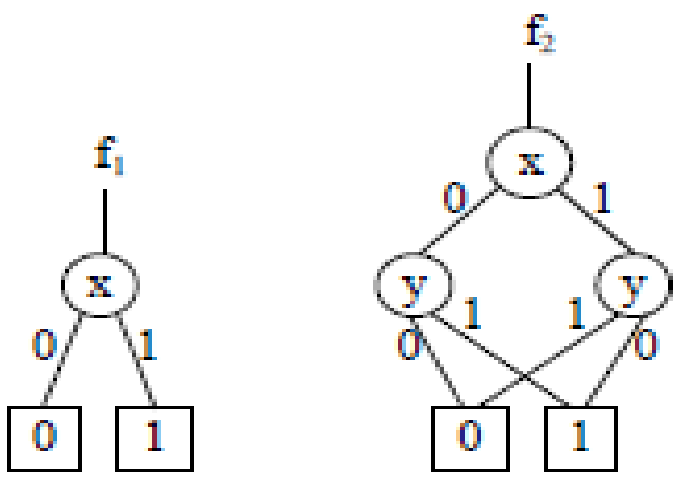

(d) ROBDD

0 1 0 0 0 0 0 0
0 1 1 0 0 0 0 0
1 1 0 1 1 1 0 0

(e) RM Spectrum

$$\begin{bmatrix} 0 & 1 & 0 & 0 & 0 & 0 & 0 & 0 \\ 1 & 0 & 0 & 0 & 0 & 0 & 0 & 0 \\ 0 & 0 & 0 & 1 & 0 & 0 & 0 & 0 \\ 0 & 0 & 1 & 0 & 0 & 0 & 0 & 0 \\ 0 & 0 & 0 & 0 & 0 & 0 & 1 & 0 \\ 0 & 0 & 0 & 0 & 1 & 0 & 0 & 0 \\ 0 & 0 & 0 & 0 & 0 & 1 & 0 & 0 \\ 0 & 0 & 0 & 0 & 0 & 0 & 0 & 1 \end{bmatrix}$$

(f) Matrix Representation

Figure 2.14 Different Circuit Design Entry Formats

**RM-Spectrum** is basically obtained form the Positive Polarity Reed–Muller Expansion of the given circuits, in-fact a RM-Spectrum represent the Positive Polarity Read-Muller Expansion (PPRM) in other way but it is used in some synthesis method. RM-Spectrum can also be used in transformation based synthesis approach[22].

## 2.3 Reversible Circuit Design

Toffoli, Fredkin and Peres have given their reversible gates in 1980's, these gates are used to implement the Boolean functions. Network of reversible gates to implement the specific Boolean function is called reversible circuits. Formally a combinational reversible circuit is an acyclic combinational logic circuit in which all gates are reversible, and interconnected without explicit fanouts and loops.

### 2.3.1 Reversible Libraries :

Reversible Libarary is collections of reversible gates, for a library L, an L-circuit is composed only from gates of L, and that circuit is called L-constructive circuit, when a library consist of a single gate(gate type) we use the gate name instead of L.

- **NCT-Library:** Tommaso Toffoli proposed a generic NCT-circuit constrcution for an arbitrary reversible function, circuit implementing these function called NCT-constructible circuit. NCT-Library is also stanardized by Toffoli in 1980[9].
- **NCTS-Library:** Fredkin and Toffoli presents a new gate Fredkin gate, Several cases of the generalized Fredkin gates can be found in the literature. A gate with no controls, FRE(x1,x2), is usally called SWAP since it swaps the signals on x1 and x2. Swap gate is added to NCT library and new library standarized with the name NCTS in 1982[10].
- **NCTPS-Library:** In 1985 Newly proposed Peres gate is also added to this library and library is called NCTPS.
- **NCTSFP Library:** In 2005 NCT library Swap,Fredkin and Peres gates are added and new library is formulated called NCTSFP.

### 2.3.2 CAD Flow for Reversible Circuit Design :

To develop the circuits, a number of different synthesis approaches are proposed by different authors. The general flow of design of reversible circuit follows the steps as shown in figure 2.11 [25].

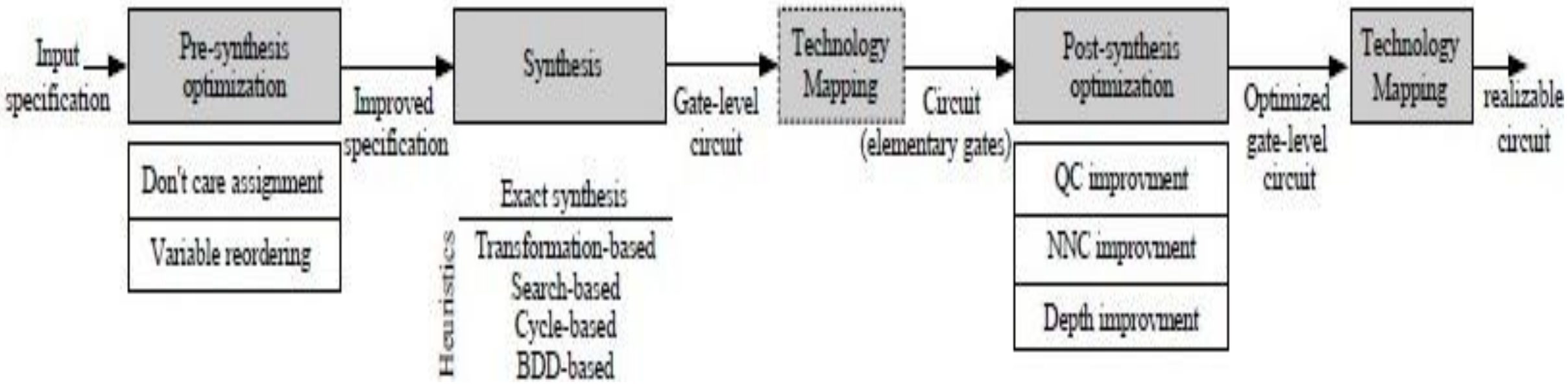

Figure 2.15 General Flow Used in the Reversible Logic Synthesis Methods

#### 2.3.2.1 Pre-Synhtesis Optimization:

To implement an irreversible specification using reversible gates, Ancillae should be added to the original specification, this steps deals with the optimization of number of Ancillae, garbage and ordering of output line. This process can be either perfromed prior to synthesis or in a unified approach during synthesis[25].

#### 2.3.2.2 Synthesis

Synthesis means seeking reversible circuits that satisfy a reversible specification. Many synthesis methods are given in literature like Transformation based, Cycle based, BDD based, Search based, Programming language based etc. Different authors with different time proposed new mehtod of synthesis and improvement in these method of synthesis.

- **Transformation Based Method:** Iteratively select a gate so as to make a function's truth table or RM Spectrum more similar to the identity function. These methods are divided into two parts. First is called Embedding of irreversible function and second step is synthesis using transformation. This method was proposed by D**. Michael Miller** in **2003**[26].
- **Cycle Based Method:** This method decomposes a given permutation into a set of disjoint(often small) cycles and synthesis individual cycles separately. This method was proposed by **Vivek V. Shende** in **2003**[27,28].
- **Search Based Method:** This method traverses a search tree to find a reasonably good circuit. These methods mainly use the PPRM expansion to represent a reversible function. The efficiency of these methods is highly dependent on the number of circuit lines and the number of gates in the final circuit. This method was proposed by **P. Gupta** in **2006**[29].
- **BDD- Based Methods:** This method uses binary decisions digrams to improve sharing between controls of reversivble gates. These techniques scale better than others. However, they require a large number of ancillae qubits(a valuable resource in fledgling quantum computers). This method was proposed by **Robert Wille** and **Rolf Drechsler** in **2009**[30].
- **Programming Language Based Synthesis :** The approaches disussed above rely on Boolean descriptions, thus do not allow the design of complex reversible systems. Consequently, higher levels of abstaractions have been considered leading to the development of hardware description languages. Hardware descripition langauges like Syrec are in development process for their efficient use. This method was proposed by **Robert Wille**, **Sebastian Offermann** and **Rolf Drechler** in 2011[31].

#### 2.3.2.3 Post – Synthesis Optimization

In post synthesis optimization are often sub-optimal. Some improvements can be achieved by local optimization. Some post-synthesis optimization are as follows:

- **Improving Gate Count and Quantum Cost:** To improve the quantum cost of a circuit, two techniques attempt to improve individual sub-circuits one at a time. Offline synthesis techniques is proposed which user per-computed tables and in year 2010 online synthesis technique is proposed which is of dynamic nature. This method was proposed by **Aditya. K. Prasad**, and **Dmitri Maslov** in **2006**[32].
- **Improving Locality:** For the implementation of a given compuation on a quantum architecture with restricted qubit interactions, one may use SWAP gates to move gate qubits towards each other as much as required. A generic approach can also be used to either reduce the number of SWAP gates or find the minimal number of SWAP gates for a circuit. This method was proposed by **Kutin** and **Takahashi** in **2007**[33].
- **Reducing Circuit Depth [2008] :** To realize a low-depth implementation of a given function, consecutive elementry gates with disjoint sets of control and target lines should be used to provide the possibility of parallel gate execution. This method was proposed by **Dmitri Maslov** in **2008**[34].

## 2.4 Tools

Reversible circuits become popular need of software tools have been felt for simulating reversible circuits. Development work for such tools is in progress, few tools have been proposed but not standardized. The following tools as follows:

- **Rev Kit :** It is a open source tool, it accepts reversible circuit in PLA form and schematics of circuit can be seen with some operations like equivalance checking, joining, calculating quantum cost can be perfromed.
- **RC Viewer:** It is circuit viewer tool, accepts circuit in *.real from, improved version have introduced with the name of RC Viewer +.
- **RC Dev :** It is developed by Nitin Purohit and Dr. S.C. Jain. In this tool we have the function like NCTPFS library based circuits, inter-format conversion and logic synthesis through different design entry formats.

- **RC Test :** It is developed by Anugrah Jain and Dr. S.C. Jain. In this tool we have function like Parity Preservation of the reversible circuit, Generation of the reversible circuit from ESOP syntheis and convert a parity preserving reversible circuit into an online testable reversible circuit.

The available tools are not standardized and have limited functionality, these tools are not user friendly for developing reversible circuit.

## 2.5 Others

Reversible computation is emerging technology, so the work in following different areas of this fileld are in progress but not at standarized state.

### 2.5.1 Reversible Programmable Gate Array [RPGA]

Reversible computing is emerging technology and interest towards reversible circuits, hence need of a regular structure felt which should be counterpart of Irreversible Field Programmable Gate Array, in year 2011 M. Perkowaski and P. Kerntopf presented design of Reversible Programmable Gate Array (RPGA) based on regular structure to realize binary functions in reversible logics[35]. Structure is based 2*2 net structure in which Arbitrary Symmetric function can be realized in a net without repeated variables. Only tool available in RPGA is RPGA Sim.

#### 2.5.1.1 Tool for RPGA

- **RPGA SIM** : It is developed by Pankaj Israni and Dr. S.C. Jain. In this tool we have function like symmetry analyzer for reversible circuit, generate RPGA structure of any given input, and generate response for any symmetric circuit.

### 2.5.2 Reversible Programming Language

Reversible computing have edge over conventional systems, there is need of programming language which can be used in such systems.

- **Janus:** In 2007 R Gluck and T. Yokoyoma proposed a reversible programming language and its self invertible interpreter[36].
- **Syrec:** Robert wille, Sebastian Offermann, and Rolf Drechsler proposed a new programming language in 2011. Syrec is basically a hardware description language which is based on previously presented reversible language "Janus", this programming language allow to specify and afterward to automatically synthesize reversible circuits [37].

## 2.6 Survey Extraction

In the above literature survey it is observed that Reversible circuit offer a great advantage over Conventional circuit in design size, speed, cost and time. But implementation platform available is only conventional, which does not support reversible circuits.

A number of design tools and circuits are available for conventional Irreversible designs, but the same can not be used or developed in reversible logic because of paucity of tools. In order to use the above tools and circuit, we attempt to develop a tool named "IRC2RC" that convert Irreversible circuits into Reversible Circuits.

# Chapter 3

# CONVERSION APPROACH AND DESIGN ALGORITHMS

This chapter elaborates conversion issues and apporach for mapping irreversible circuit to reversible circuit. Irreversible circuits do not have same number of bits in input and output. Also the gates do not have adequate information to reconstruct the input. Hence, the issues and their solution is described in the following sections.

## 3.1 Conversion Issues

The conversion has to provide adequate provision in input as well as output to make reversibility possible. Additional input(s) and output(s) are required to incorporate reversibility. The related issues and their proposed solutions are described in this section.

### 3.1.1 Garbage Output and Costant Input

Every irreversible gate has one or more inputs and one output only. The output has insufficient information to reconstruct input. In reversible circuit the required output is known as target line and additional lines required to incorporate reversibility known as garbage output.

However sometimes additional lines are required at input also they are known as constant input. The issue is to obtain adequate number of garbage as well as constant input lines for conversion. Figure 3.1 shows an irreversible AND gate converted into equivalent reversible gate.

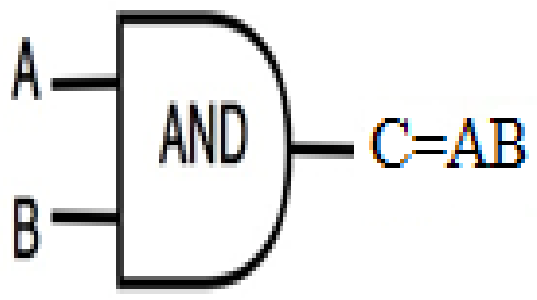

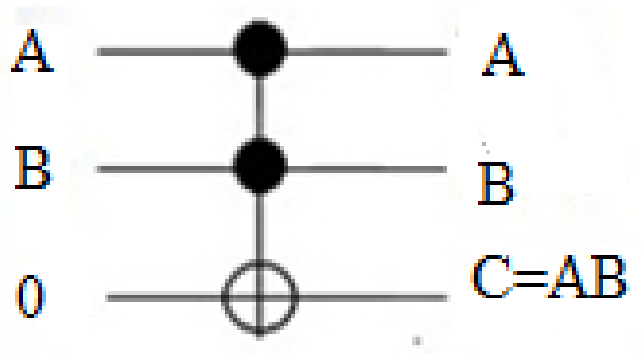

Figure 3.1 (a) AND Gate (b) Equivalent Reversible AND Gate

In order to obtain equivalent reversible gate 3*3 Toffoli gate is used where two input namely A and B are the same and output is available at C. A constant input zero is required at input in addition to A and B. Ofcourse A and B are also available at output as garbage values. The constant input and garbage outputs are not required as per the circuit output but required for reversibility.

### 3.1.2 Mapping Library

Since the irreversible gate are no more used, equivalent reversible gates require mapping of irreversible gate to equivalent reversible gate. This will constitute a reversible mapping library. The following equivalent gate library has been developed for commonly used seven irreversible gates. The figure 3.2 (a-g) shows the equivalent reversible gates for a mapping.

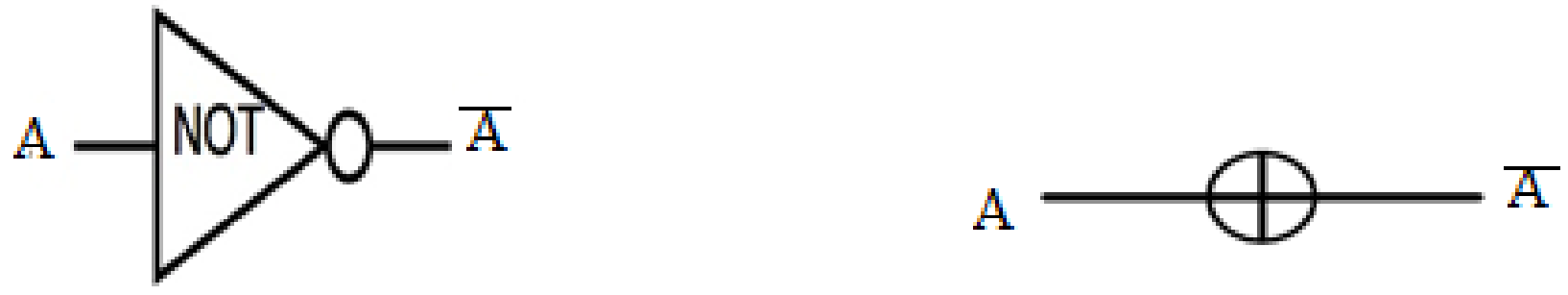

Figure 3.2 (a) Irreversible NOT Gate and its Equivalent Reversible NOT Gate

As shown in figure 3.2 (a) NOT gate by itself is reversible and hence can be implemented using 1*1 reversible Not gate.

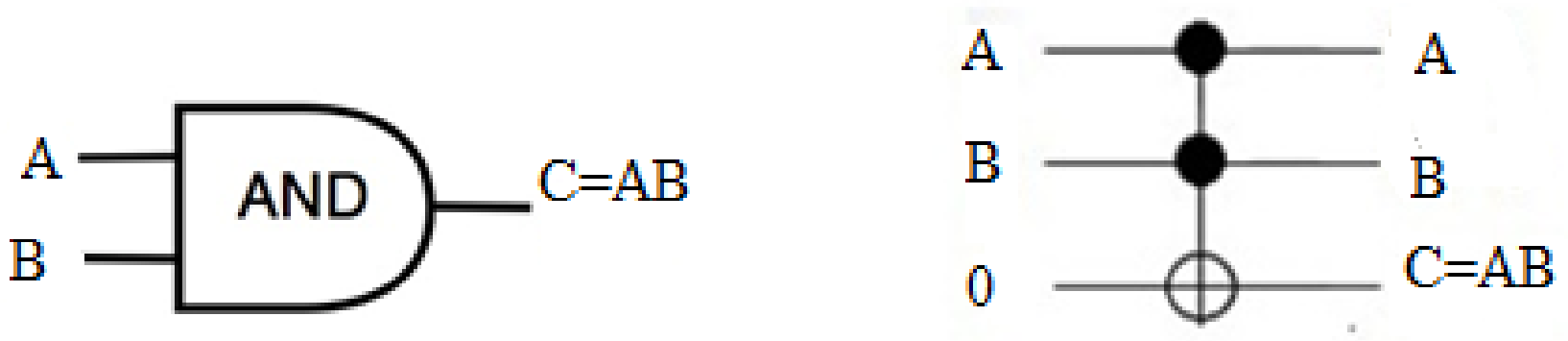

Figure 3.2 (b) Irreversible AND Gate and its Equivalent Reversible AND Gate

Figure 3.2 (b) shows an implementation of 3*3 Toffoli gate in which third input is costant ‘0’ input. A and B inputs are produced as output which are garbage output. Primary output of the gate is shown C=AB.

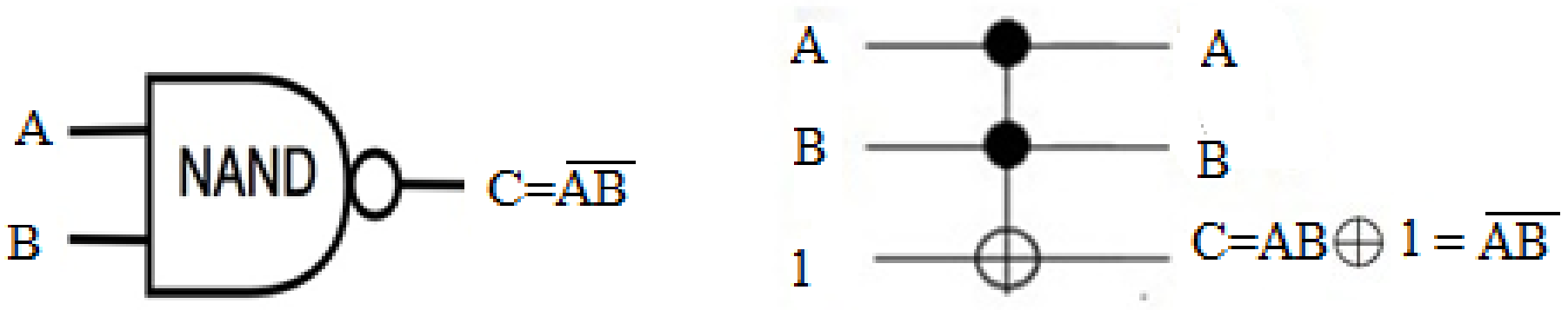

Figure 3.2 (c) Irreversible AND Gate and its Equivalent Reversible NAND Gate

Figure 3.2 (c) shows an implementation of 3*3 Toffoli gate in which third input is costant ‘1’ input. A and B inputs are produced as output which are garbage output. Primary output of the gate is shown C=$\overline{AB}$.

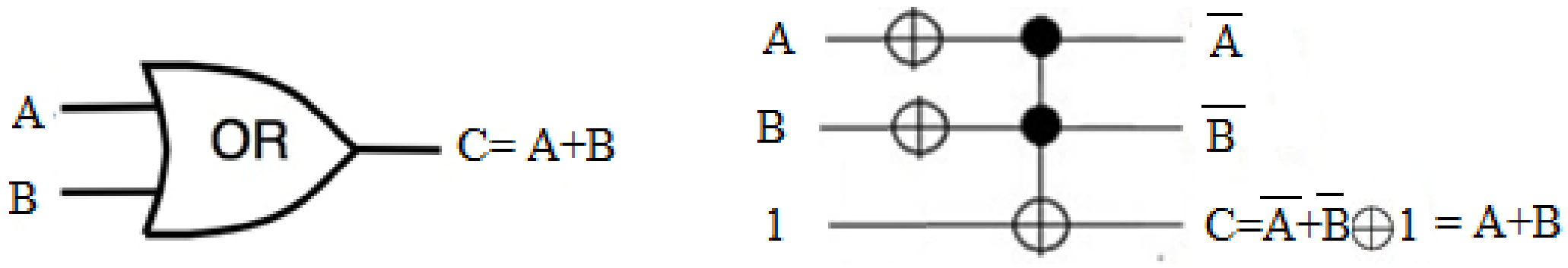

Figure 3.2 (d) Irreversible OR Gate and its Equivalent Reversible OR Gate

Figure 3.2 (d) shows an implementation of 3*3 Toffoli gate with two NOT gate in which third input is costant ‘1’ input. A and B inputs are produced as output which are garbage output. Primary output of the gate is shown C=A+B.

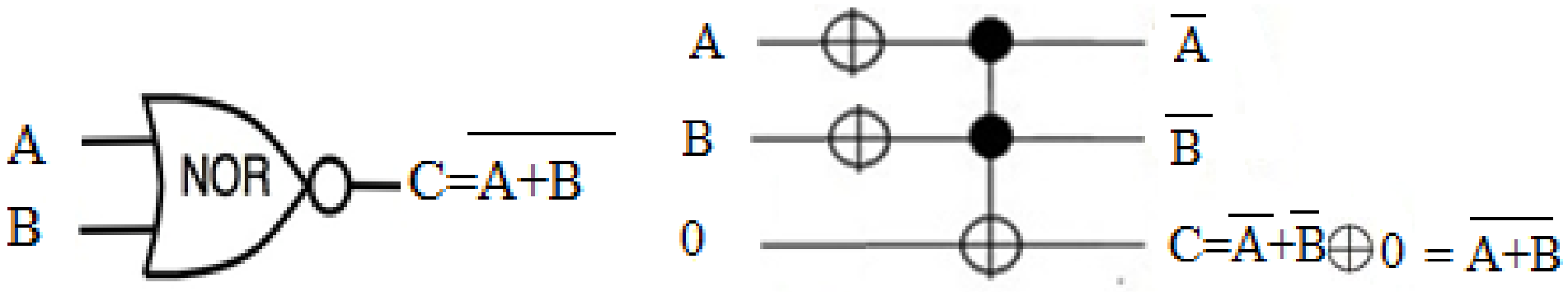

Figure 3.2 (e) Irreversible NOR gate and its equivalent reversible NOR gate

Figure 3.2 (e) shows an implementation of 3*3 Toffoli gate with two NOT gate in which third input is costant ‘0’ input. A and B inputs are produced as output which are garbage output. Primary output of the gate is shown C=$\overline{A+B}$.

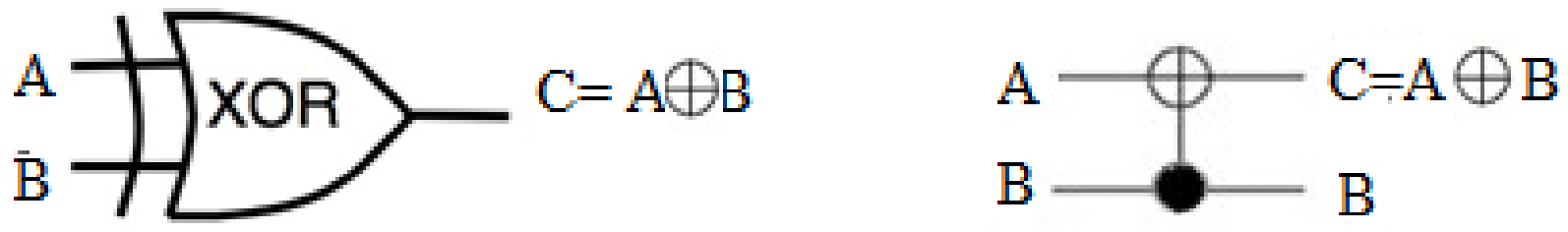

Figure 3.2 (f) Irreversible XOR gate and its Equivalent Reversible XOR Gate

Figure 3.2 (f) shows an implementation of 2*2 Feynman gate. A and B inputs are produced as output in which B is garbage output. Primary output of the gate is shown C=A⊕B.

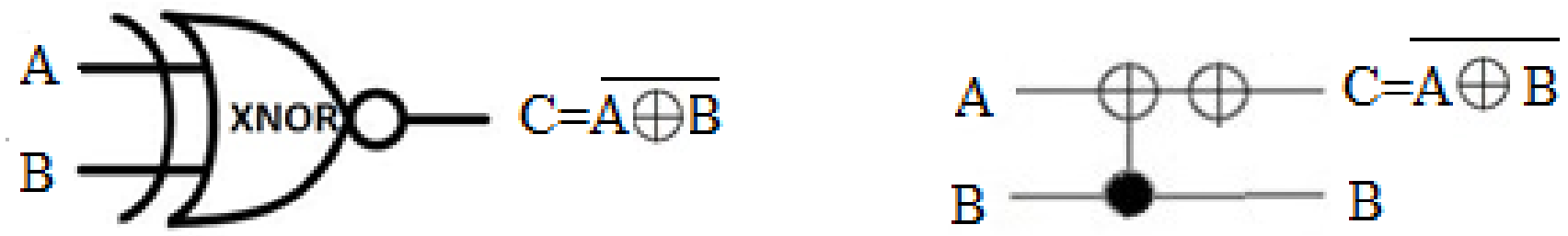

Figure 3.2 (g) Irreversible XNOR Gate and its Equivalent Reversible XNOR Gate

Figure 3.2 (g) shows an implementation of 2*2 Feynman gate with NOT gate. A and B inputs are produced as output in which B is garbage output. Primary output of the gate is shown C=$\overline{A \oplus B}$. Figure 3.2 (a-g) shows a conversion of irreversibe gate to its equivalent reversible gate.

### 3.1.3 Equal Number of Input and Output

The reversible circuit so generated should have equal number of input and output. Intermediate gates having additional inputs and outputs must start from primary inputs and end at primary outputs. Figure 3.3 (a-c) shows a conversion of circuit to equivalent reversible circuit.

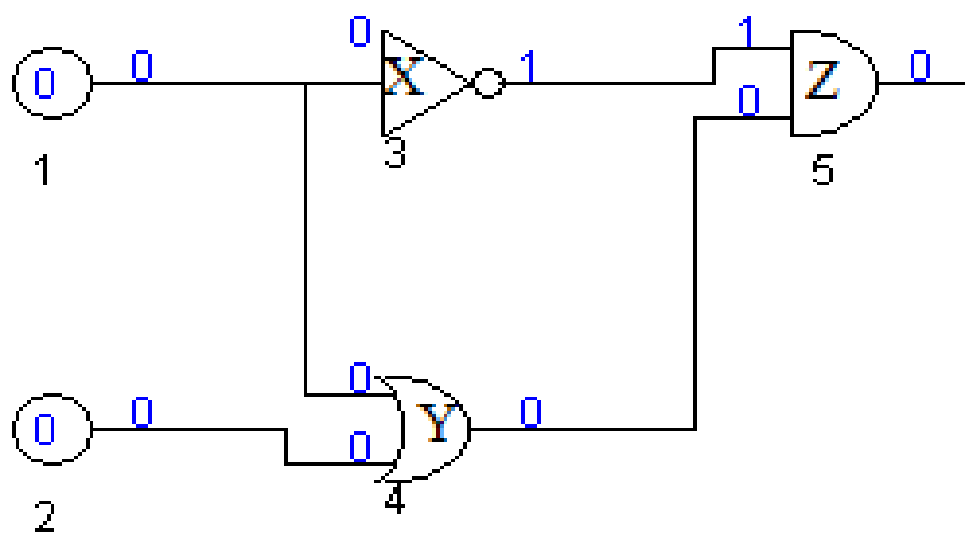

Figure 3.3(a) Irreversible Circuit

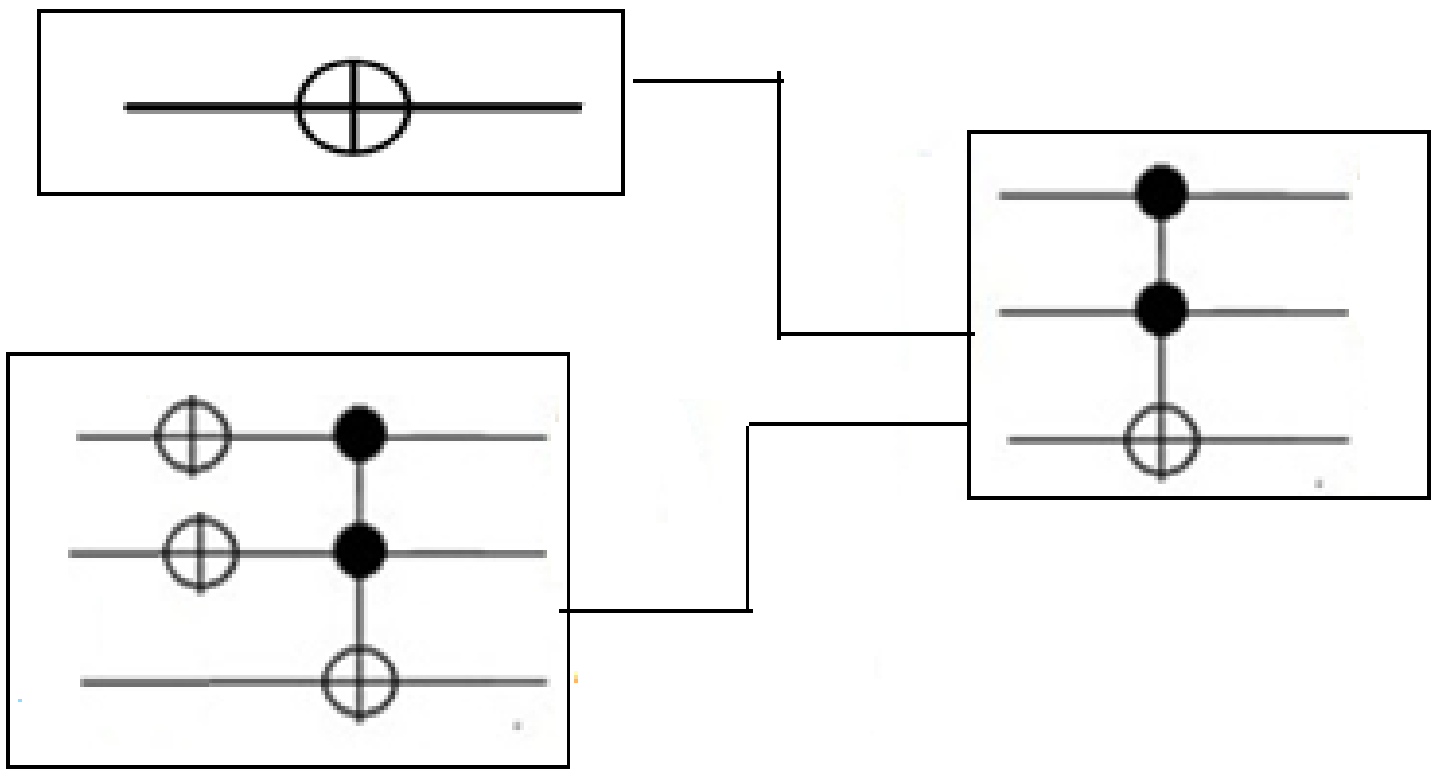

Figure 3.3 (b) Replacement Gate Based Circuit

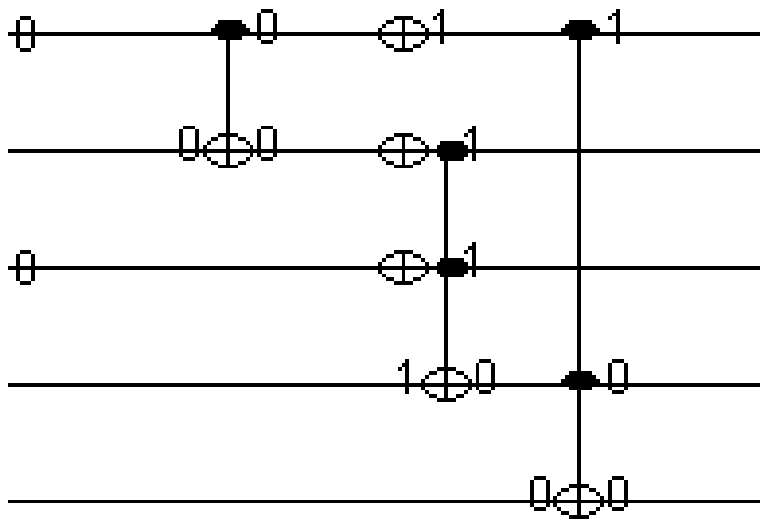

Figure 3.3 (c) Converted Reversible Circuit

Figure 3.3 (a) shows a irreversible circuit. While conversion of gates Y and Z require additional constant input and additional garbage output, which are represented in figure 3.3(b). The same lines are to be extended to primary input to primary output for complete conversion which is shown in figure 3.3 (c). The algorithms and procedures to be implemented are detailed in subsequent sections.

## 3.2 Apporach for Conversion

There are two apporaches identified for conversion namely Truth table based approach and Replacement based apporach. Former apporach is based on generating a truth table for the given circuit and then using reversible synthesis tools to generate reversible circuit. Later apporach is based on with reversible counterpart from mapping library. Figure 3.4 shows both the approaches.

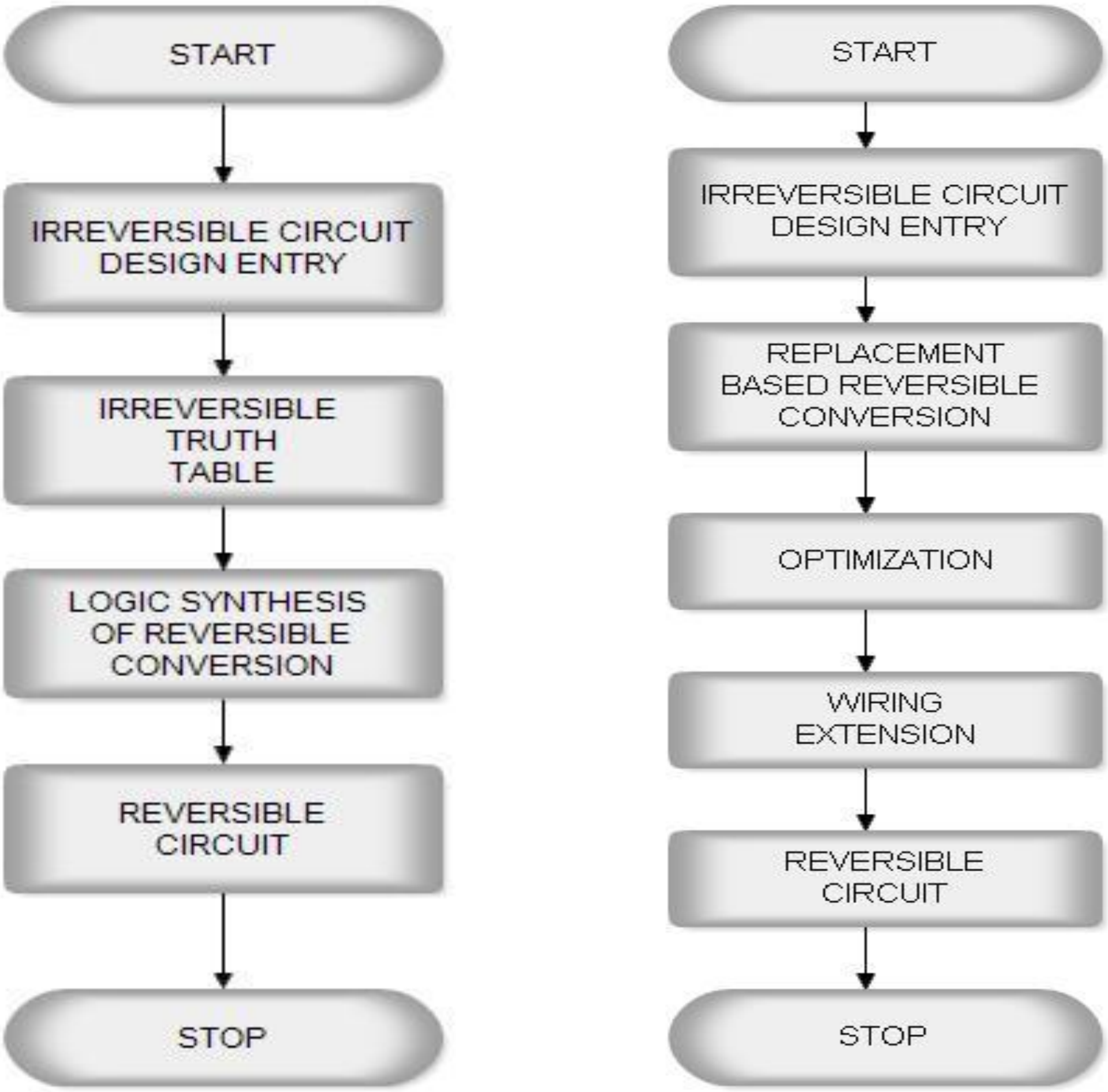

Figure 3.4 (a) Truth Table Based Approach (b) Replacement Based Approach

In the former approach irreversible circuits are converted into irreversible truth table followed by conversion into reversible circuit using synthesis tool. This approach requires reversible synthesis tools which are in scarcity and does not have any advantage of using existing circuits. Hence this approach is not considered in our project.

In the later apporach, the irreversible circuits can be converted directly into reversible circuits which gives a advantage that reversible synthesis tools are not required. This approach truely opens a door for irreversible circuits to be directly used after conversion. Of course design entry of irreversible circuits in schematic format is necessary in both the approaches.

## 3.3 Algorithms

As we know that reversible circuit do not allow feedback and fanout, handling of feedback and fanout issues of irreversible circuits is one of the tedious task. To deal with this we have developed two algorithms. First algorithm processes the circuits having no feedback and no fanout and second algorithm works as for preprocessor for the circuits having fanout. The preprocessor algorithm removes feedback and writes the circuit in intermediate form.

### 3.3.1 Irreversible Circuit Design Entry Algorithms

The first step shown in figure 3.4 (b) and it is irreversible design entry. The irreversible circuit is entered in schematic format. Schematic is divided into appropriate time slots and a irreversible gate library is used for design entry. Figure 3.5 shows the irreversible circuit design entry procedure.

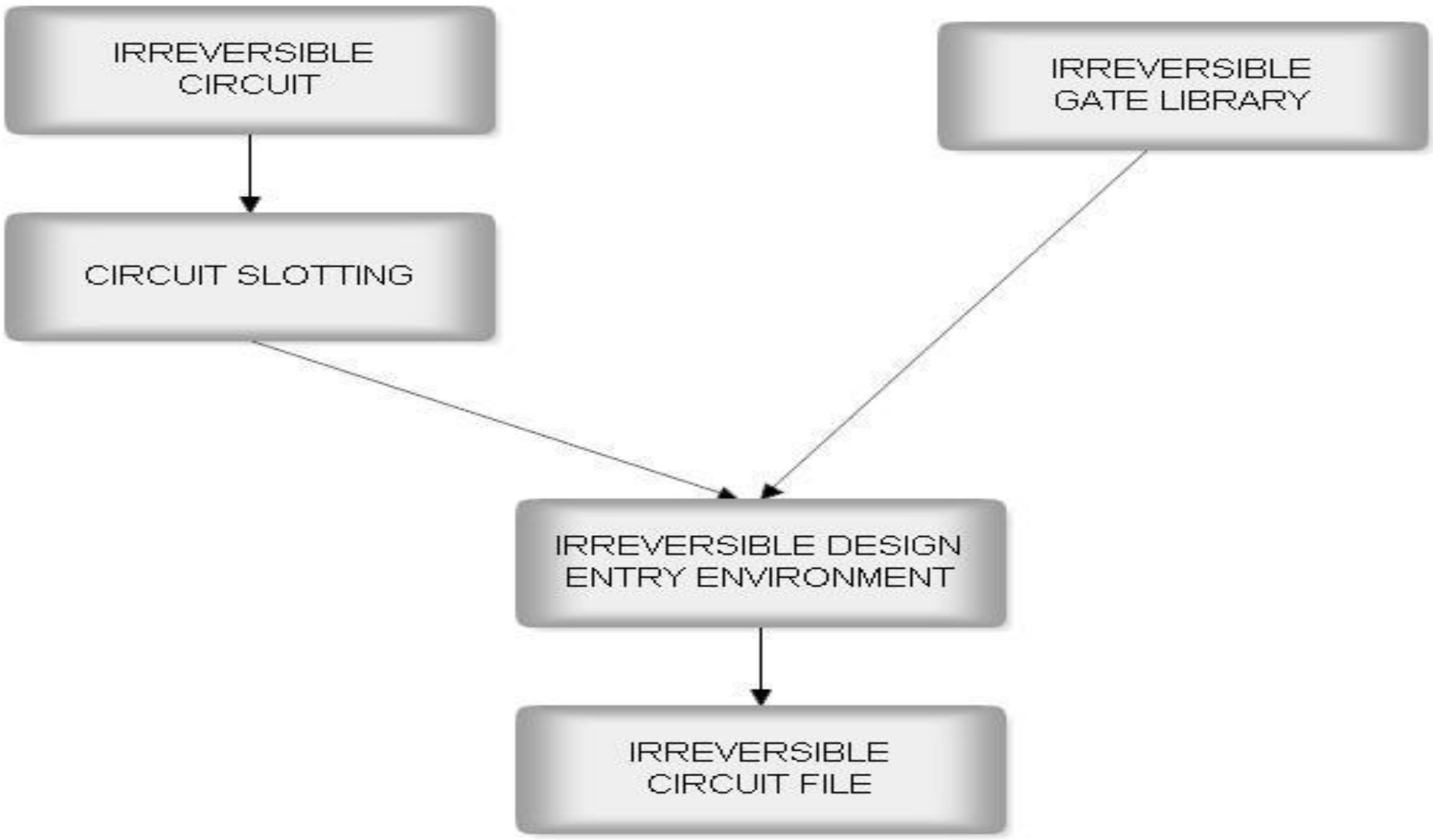

Figure 3.5 Irreversible Circuit Design Entry Procedure

Since the conversion take place from input side to output side, it is required to convert any gate only if predecessor gate have been converted. For this purpose circuit slotting of the circuits are needed so that all gates in previous slots must be converted into corresponding reversible gates before current slot conversion.

At present the slotting process is done manually and designs are entered through drag and drop facility from irreversible library. Storage and reterival facility of irreversible circuit is also implemented. However circuit slotting and preprocessor algorithm have been developed and explained in subsequent subsections.

### 3.3.2 Circuit Slotting

The circuit is divided into time slots so that slot by slot design entry as well as conversion can be facilitated. The circuit slotting can be done only for irreversible circuits having no fanout and no feedback. The algorithm for the slotting is described below.

---

**Algorithm 3.1 Slotting Algorithm**

---

**/*** Assignment of slot no. to every gate and nets will be done in this algorithm starting from primary input set till primary output set. */

/* Each slot will contain one gate set and one net set. Which means the set of gates and set of nets are part of that slot.*/

**/*** The circuit format is considered as BLIF or similar format. Which means the format describes primary input set (net names), primary output set (net names) and gate names with connected net names in input and output.***/**

**/*** Slot no. zero will contain only primary input and other slot no's will contain gate followed by net.*/

**/*** For every slot, there will be a net list and a gate list passing through this slot.*/

Input : Unslotted Circuit

Output : Slotted Circuit

Description:

| | |
|---|---|
| IS : Input primary net set | OS : Primary output net set |
| CN : Current net set | CG : Current gate set |
| SN : Selected net set | SG : Selected gate set |
| SL : Slot number | NSL : Net set in current slot |
| GSL : Gate set in current slot | |

1. Initialize SL=0
2. Assign CN=IS
3. Assign SL.NSL= CN
4. For (SL=1; CN!=OS: SL++)
5. SG= set of all gates whose all input nets are in NSL of previous slot
6. Assign SL.GSL=SG
7. Remove all nets in circuit used as input in SG
8. Add output nets of SG in CN
9. Assign SL.NSL=CN
10. Continue

The above algorithm start conversion from primary input side and by selecting gates that have all input nets are in prevoius slot. This algorithm is being explained through example in step by step manner.

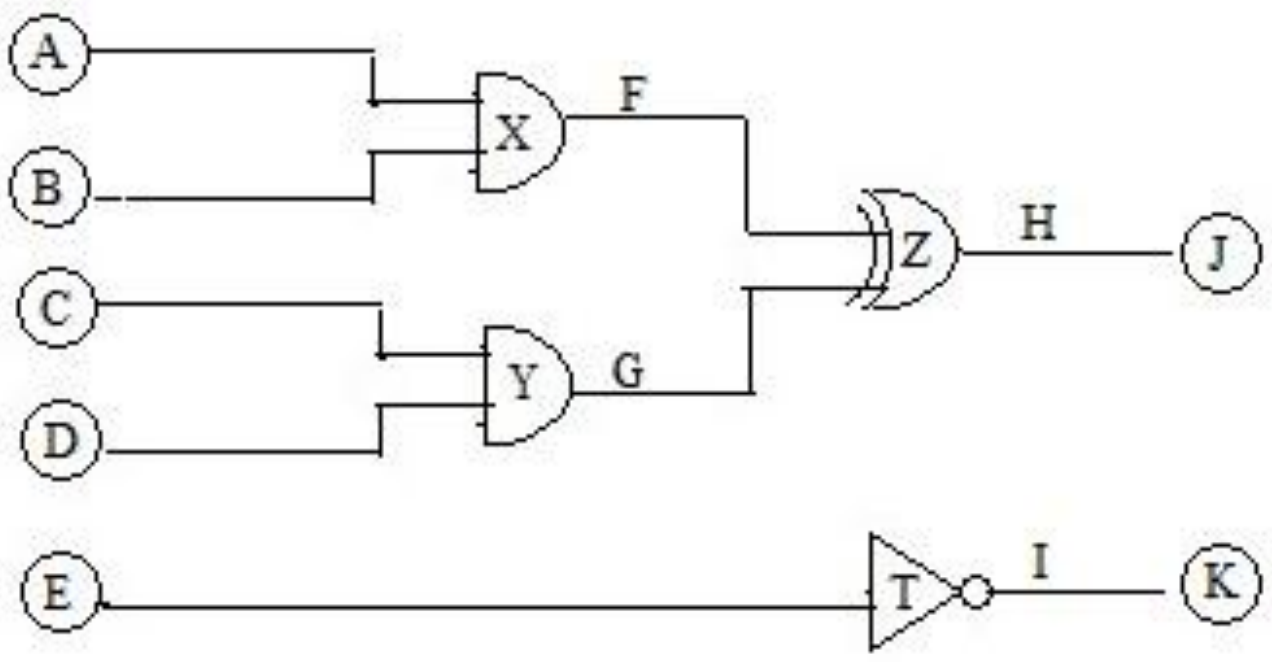

Figure 3.6 Irreversible Circuit Without Fanout

The example takes an irreversible circuit without fanout and feedback as shown in figure 3.6.

1. Slot Zero : In slot number zero all primary input net set are selected as per the line number three of the algorithm as shown in figure 3.7. These nets will be called assign nets for next slot. In figure 3.7 primary input net set contains A,B,C, D and E nets.

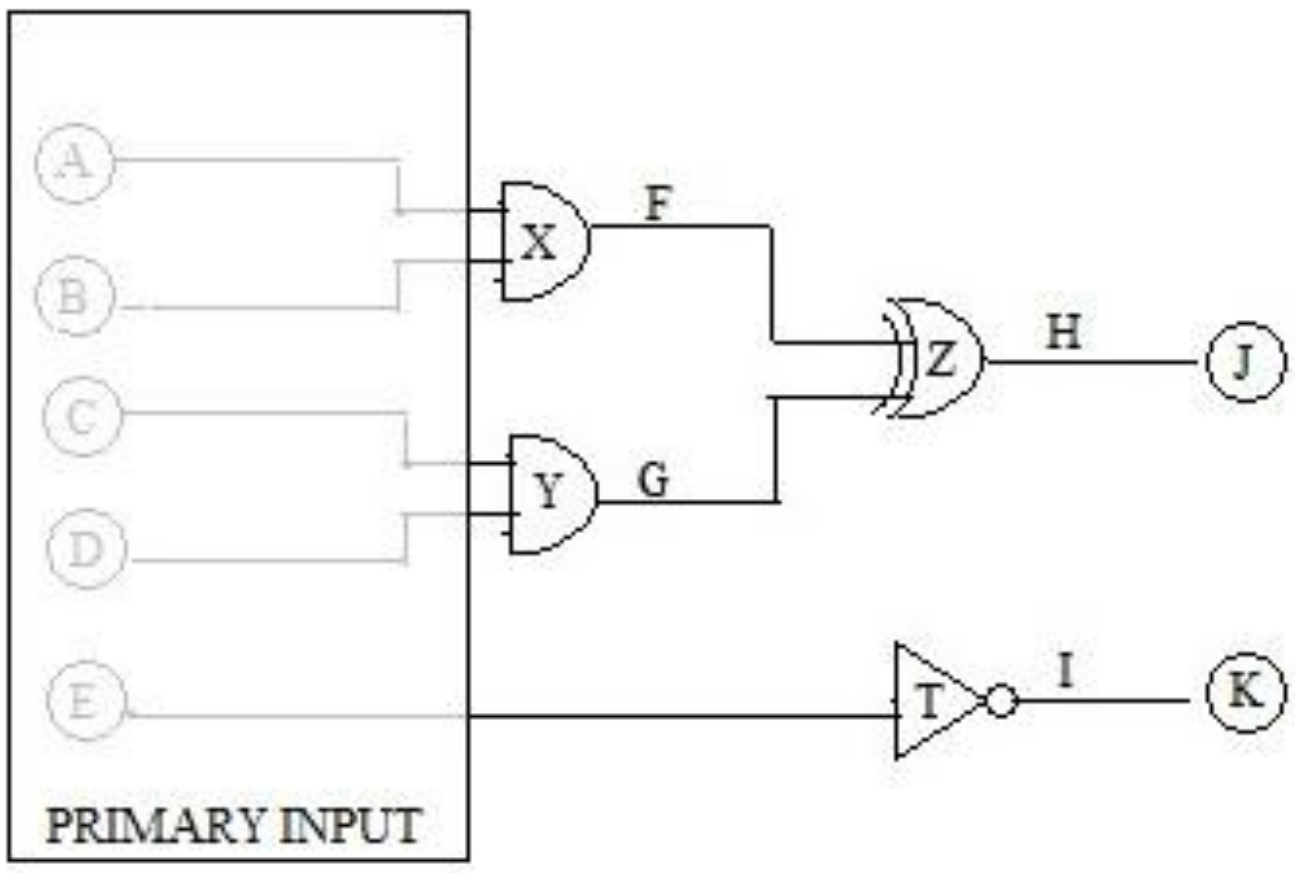

Figure 3.7 Initialize the Primary Input Net Set

2. Slot One : This is the first slot in which gate and net both will be present. The gate will be selected in line number six of first iteration. This selection will be done at line number seven of the algorithm. In figure 3.8 X and Y gates are selected in gate set where as all output nets of X and Y gates namely F, G and passing net E will be selected in current slot.

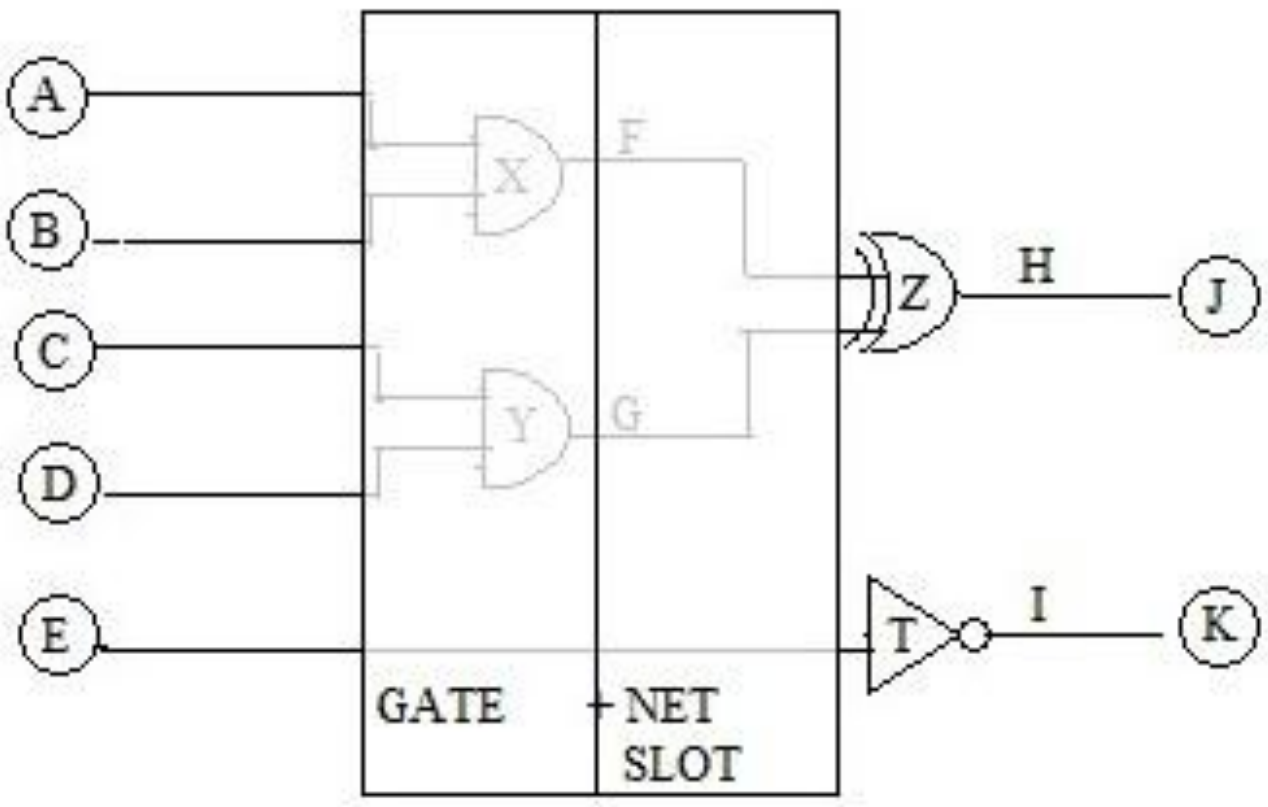

Figure 3.8 Current Gate and Net set

3. Slot Two: In this slot net and gates will be selected in next iteration in similar manner as in case of slot one. Z and T gates, H and I nets will be selected in this slot. The nets will be primary output set of the circuit.

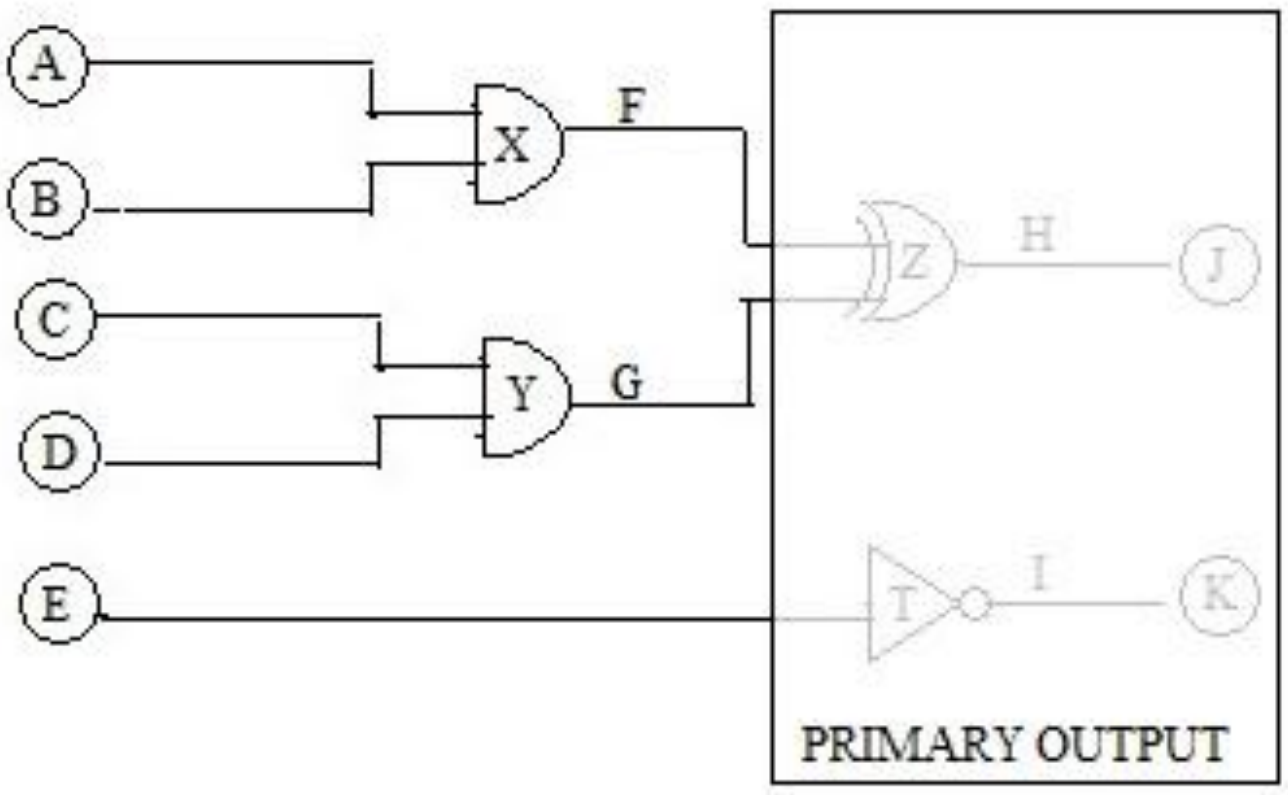

Figure 3.9 Primary Output Set

The slotted circuit stored in appropriate date structure used for conversion.

### 3.3.3 Fanout Preprocessor Algorithm

The algorihm described in prevoius subsection performs circuit slotting for irreversible circuits having no fanout. To process the circuits with fanout, the preprocessor also replaces the fanout by a Copier gate(Feynman gate of reversible circuits). Figure 3.10 shows the intermediate representation of an irreversible circuits, a Copier/Feynman gate and intermediate format after fanout processing.

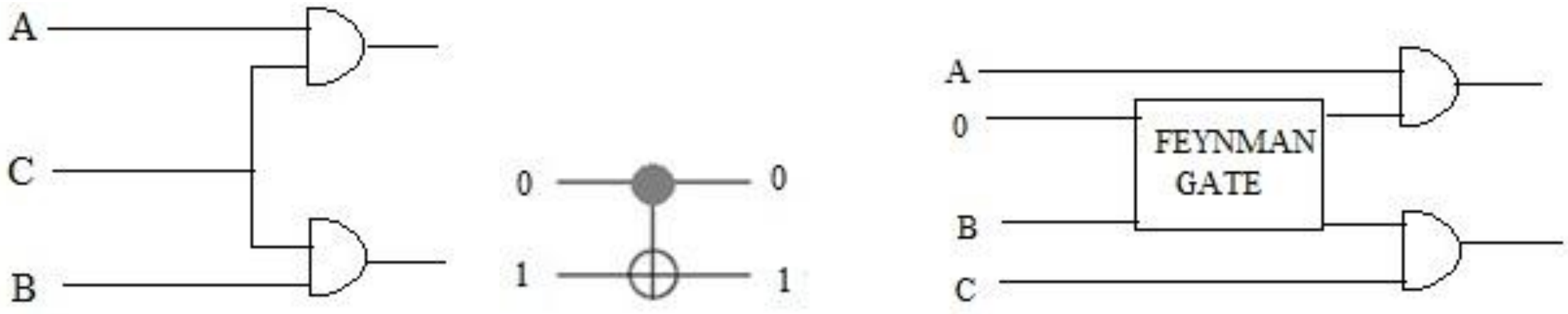

(a) Irreversible Circuit (b) Feynman Gate (c) Intermediate Format

Figure 3.10 Intermediate Representation of Irrversible Circuit

The intermediate format is used for an irreversible circuit without fanout and Feynman gate as it replaced as it is in converted reversible circuit.

The preprocessor algorithm assumes the circuit stored in BLIF like format. BLIF like format stores details of the nets as well as gates. A typical BLIF like format of AND gate is given below.

```
.model and
.inputs a b
.outputs c
.names a b c
11 1
.end
```

The above format contains model name ‘and’, primary input net name a,b and output net name c. The ‘names’ specifies a gate followed by input and output nets.

The algorithm constructs a net list of nets from the BLIF like format to identify fanout. For every fanout found a Copier/ Feynman gate is replaced in intermediate format and finally this format is used for conversion. The details of the algorithm as follows.

---

**Algorithm 3.2 : Preprocessor Algorithm**

---

/* A net in netlist contains net name, source gate name, sink gate names.*/

/* A net list NL is a collection of nets belonging to the circuit.*/

/* An intermediate format is the format that contains irreversible gates as well as Feynman/Copier gate but has no fanout. */

/* Primary input in the net list, source name wiil be PI.*/

/* Primary output in the net list, sink name will be PO.*/

/* A BLIF like format will be converted into an intermediate format to accmmodate having two output and it will be called intermediate format. */

/* A net description gate name followed by input nets and last one is output net.*/

Input : Irreversible Circuit With fanout
Output : Irreversible Circuit in Intermediate Format

1. Initialize net list NL to Nil
2. Initialize intermediate format with source format of circuit
3. Add all net names of PI in NL and add source as PI
4. Add all net names of PO in NL and add sink as PO
5. For every gate name
6. Assign gate number
7. Read all net names

8. For every net name
9. Update net list as per table 3.1
10. Add copier gate in intermediate format wherever required
11. Continue
12. Continue

Table 3.1 Net list of Intermediate Format

| **Net** | **Does not exist in net list** | **Exist in net list** | |
|---|---|---|---|
| | | **Another Sink** | **No other Sink** |
| Not Last | Add net gate as sink | Add copier gate store F(net, no.) | Add this gate as sink |
| Last | Add net gate as source | Add this gate as source | Add this gate as source |

At the end of the preprocessor algorithm the intermediate format will be generated and saved for circuit slotting and display.

### 3.3.4 Conversion Algorithm

This algorithm finally converts the schematic circuit captured in appropriate data structure and performs conversion of the circuit into target format. The data structure captures the circuit in slotted format and gate by gate conversion take place from primary input to primary output.

For replacement process mapping library is stored which available as a method for displaying corresponding reversible gate.

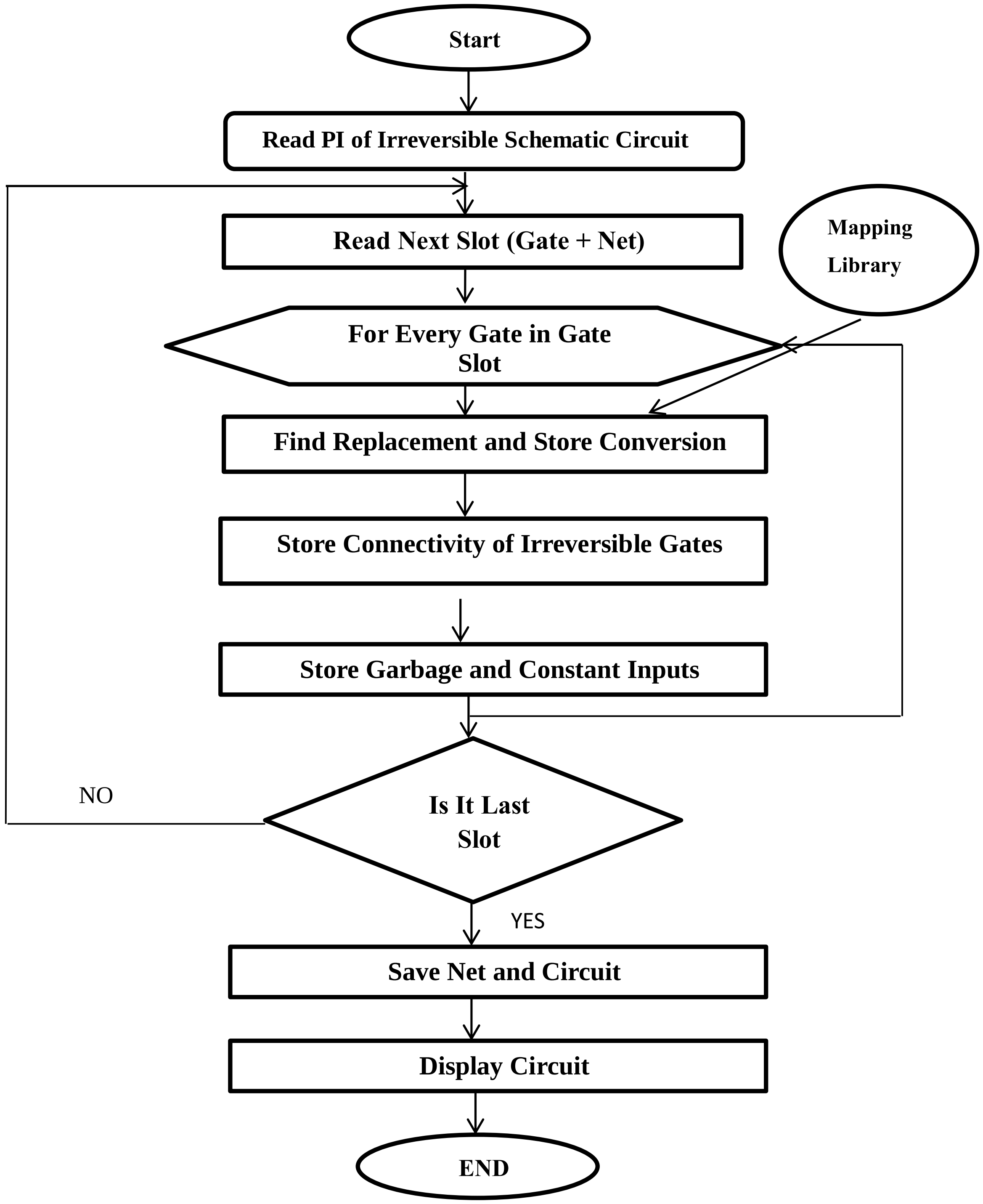

Figure 3.11 Flow Chart of Conversion Algorithm

Figure 3.11 shows conversion process. Initially slot zero primary input(PI) of the circuit is processed and then gates and corresponding nets are processed slot by slot. Each slot contains one or more gate and nets passing through or outgoing through current slot. For every gate replacement is found from mapping library and connectivity is preserved for reversible circuit display. Display module displays the conversion circuit.

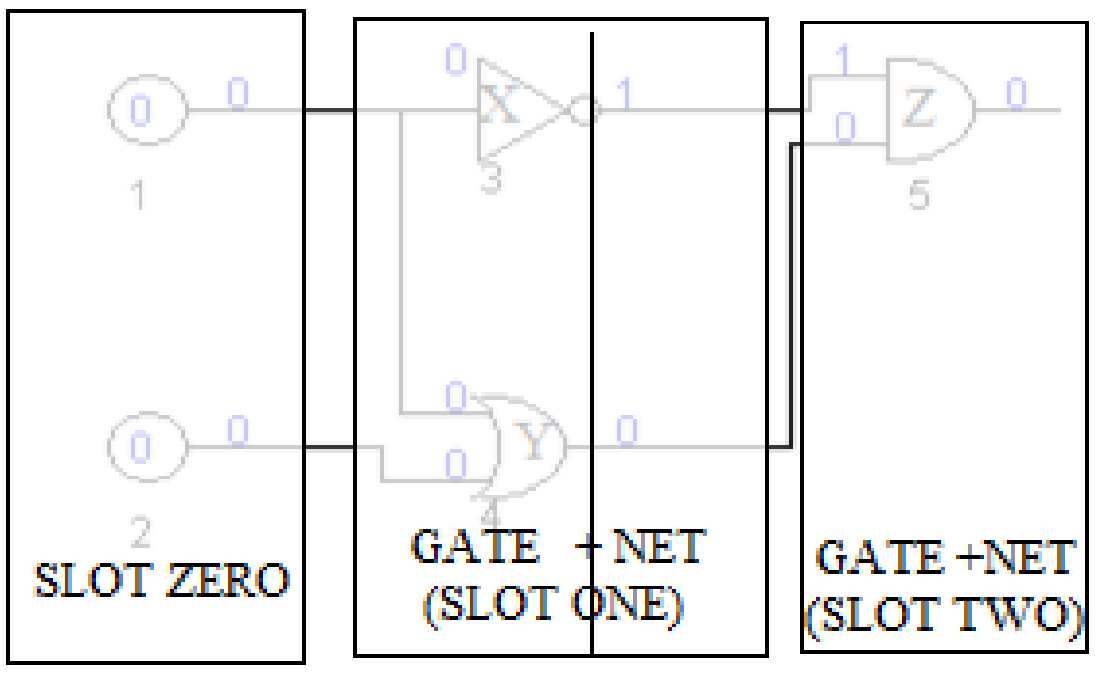

Figure 3.12(a) Irreversible Circuit Before Iteration

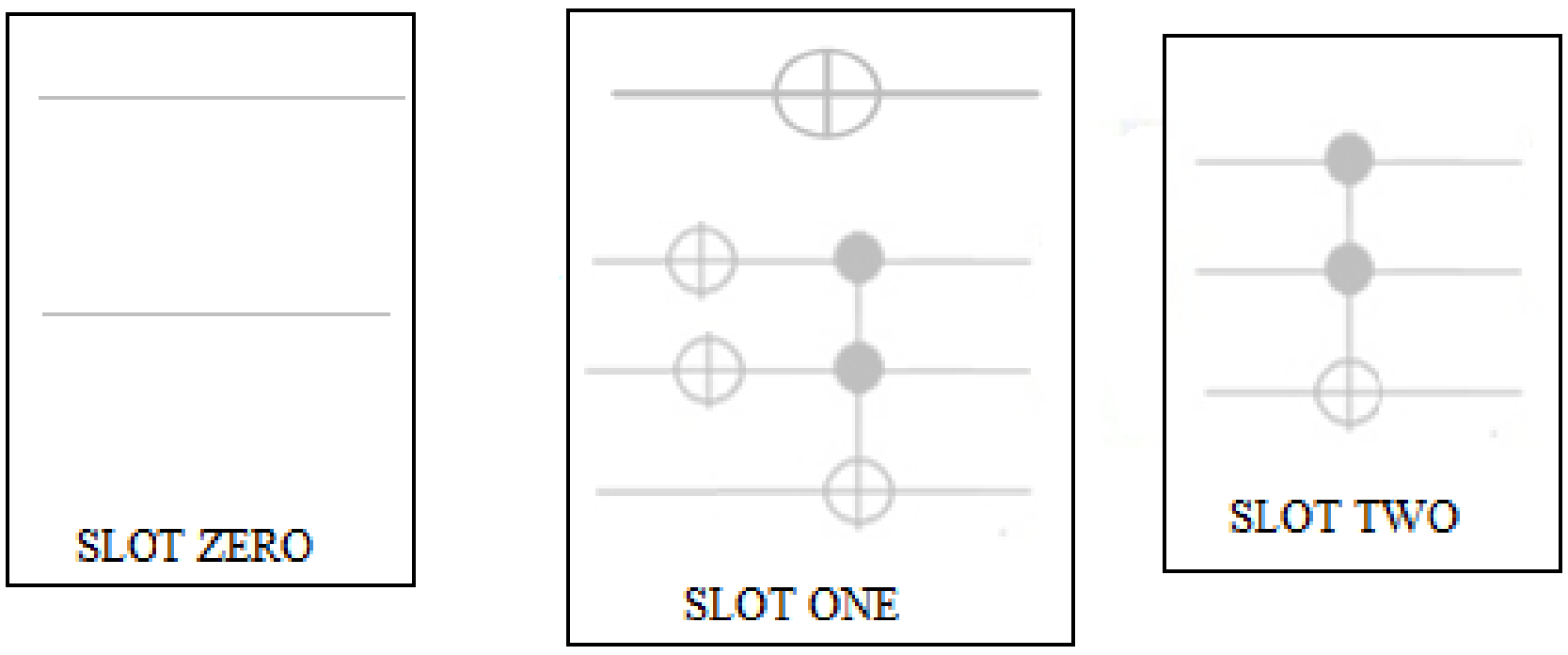

Figure 3.12 (b) Gate Replacement After Iteration

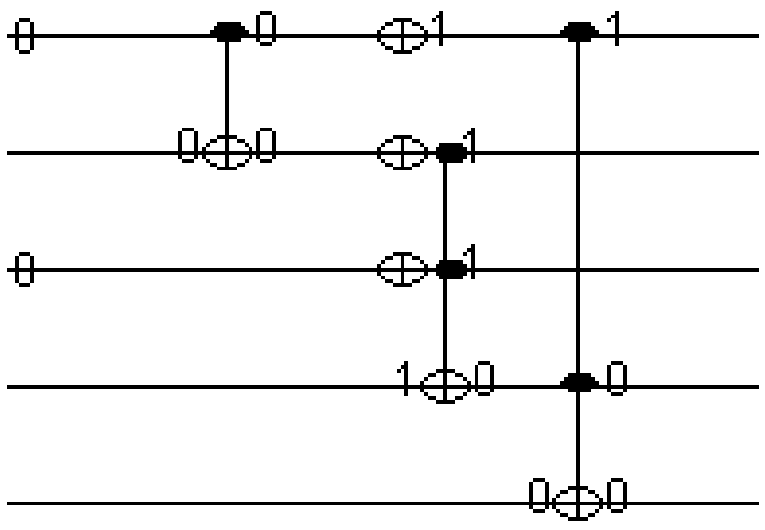

Figure 3.12 (c) Converted Circuit After Iteration

Figure 3.12 (a-c) Irreversible to Reversible Conversion

The conversion algorithm is explained the same example as shwon in figure 3.3. Figure 3.12 shows a slotted the same circuit as shown in figure 3.3 after circuit slotting.

After each iteration every slot is converted it in corresponding reversible circuit as shown in figure 3.12 (b). Figure 3.12 (c) shows the final circuit in which Corresponding/ intermediate constant input and garbage outputs are extended from primary input to primary output.

In this chapter algorithm used for conversion approach and algorithm have been described and explained through example. The conversion algorithm will be implemented in out tool named “IRC2RC” described in next chapter.

# Chapter 4

# Integration and Testing

The algorithm described in the previous chapter for conversion have been implemented in a tool named "IRC2RC". The implementation, test and verification results are described in this chapter.

## 4.1 Objective of The Tool

The tool is designed with the the following objective.

- **To Create Design Entry in Reversible Circuits:** There are two modes of design capture for irreversible circuits one is reading from file and another is to provide by through GUI facility. The entire circuit is divided into time slots from input side to facilitated slot by slot display, storage as well as for conversion. This slotting can be done by program. If the irreversible circuit read from a file in prescribed format but has to be done manually. If the circuit are entered through GUI.

  In the present project we have considered design entry through GUI and hence the slotting has been done manually. The slotting algorithm is given in previous chapter.

- **Circuit Editing and Storage Facility:** The tool is to be facilitated with partial design entry, saving, retrieval of old designs and editing current design facilities.

- **Mapping Library:** The conversion will be effected using mapping library stored in a file/methods described in language. The library is generated manually as it is one time offline job. There are seven commonly used irreversible gates for which equivalent reversible circuits have been obtained. These conversion are perfromed through methods and displayed.

- **Conversion and Drawing:** The conversion and wiring management algorithms are described in previous chapter. The algorithms are implemented and the conversion can be effected on click of a button. Post conversion wiring in reversible circuit is perfromed immediately after conversion and displayed. The implementation details are described in this sections.

## 4.2 Implementation and GUI Interface

Implementation of a tool has been done using java based GUI. The details of implementation have been described in subsection.

### 4.2.1 Graphics User Interface of the Tool

The tool provides a user friendly graphical user interface and a snapshot of the working screen is shown in figure 4.1. After making alphabet character on different portions of similar nature. The facilities converted by alphabet are described below.

**A :** (Inputs) for developing and simulating the circuit first we require the input whether it can be (0 or 1) any of them and you take multiple inputs together.

**B :** (Logic Gates) In this section known as Logic Gates, here total seven gates are showing, name are given like this AND, OR, NOT, NOR, NAND, XNOR and XOR. When we click on any of gates it will be displayed in top left corner of drawing sheet by default, now we move these gates in time lines. So we can add multiple gates together.

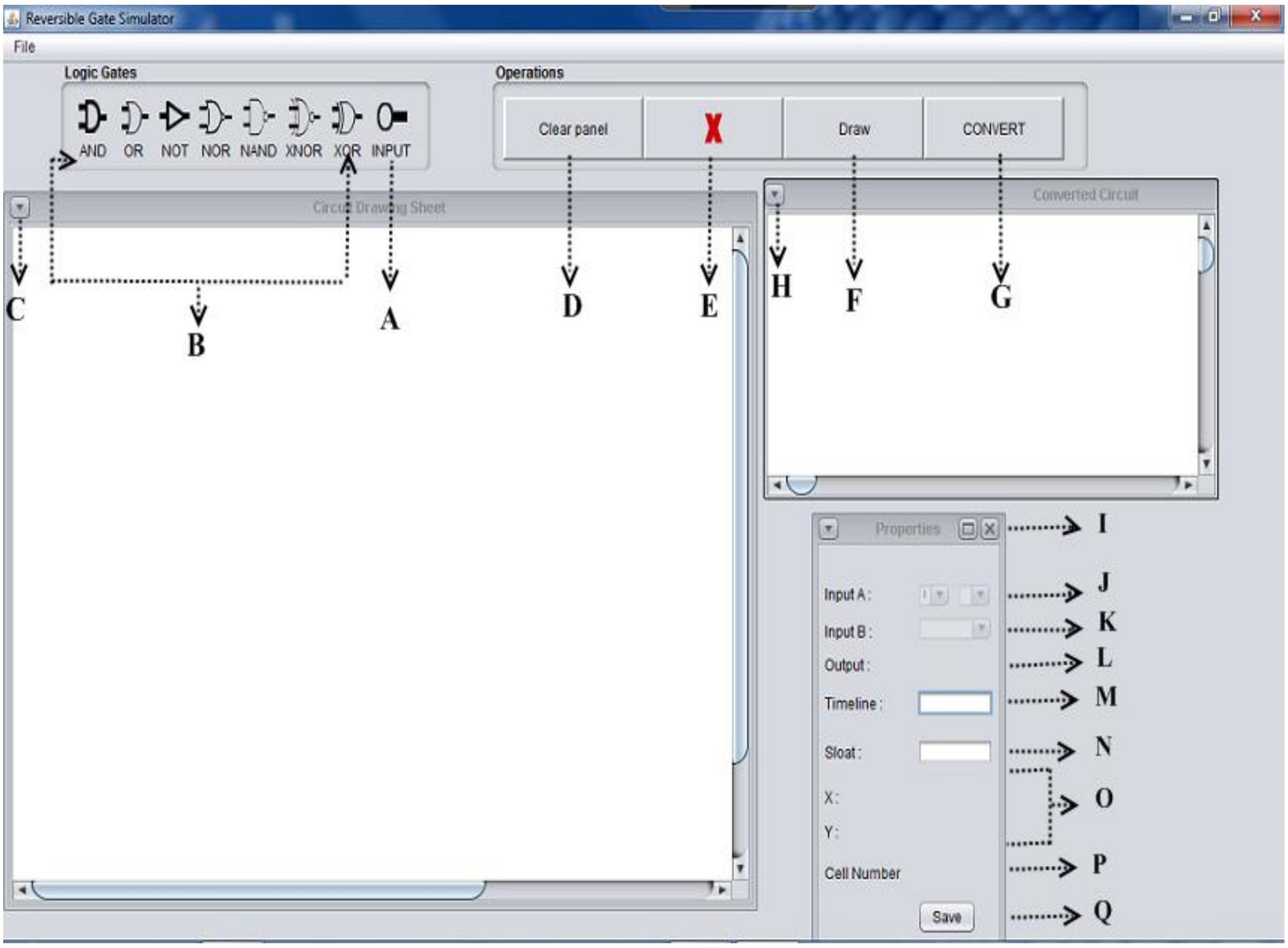

Figure 4.1 IRC2RC Tool Reperesentation

**C:** (Circuit Drawing Sheet) In circuit drawing sheet when we select the inputs then it will draw on circuit drawing sheet which is available in tool position of the tool screen. The input gate will draw in top left corner by default and then we will move it in any of the time slot in which we required. When gate is draw on top left corner then we move it, by mouse in any time slot or first time slot. We can draw multiple of gates one after another till whole circuit is drawn in this window.

**D:** (Clear Panel) This button will clear the window.

**E:** (Delete) By selecting any gate then press delete button it wiil delete the gate from data structure and help us to add other gate on their respective position.

**F:** (Draw) When we click on the draw button then on the drawing sheet wires will be connected between gates input and output according to their input references.

**G:** (Convert) Conventional circuit we will draw on the circuit drawing sheet that will convert into reversible circuit with replacement of every gate.

**H:** (Converted Circuit) This is the second display window where we will see the converted reversible circuit of binary circuit. This is our output circuit and we get conclusion here.

**I:** (Properties) In this section we have to set the property of every gates and inputs. It also give the idea that how we can set every input value of every gate and save their values in date structure. So this section have some other section like J,K,L,M,N,O,P,Q which will discuss in their subseqent terms.

**J:** (Input A) Input A requires input reference of input gates.

**K:** (Input B) Input B requires input reference of input gates.

**L:** (Output) It shows the output of gate.

**M:** (Time Line) Time line give us gate location .

**N:** (Slot) Slot location is given time line and gate location too.

**O:** (X,Y) Axis location in drawing sheet.

**P:** (Cell Number) It will give the current location of that gate.

**Q:** (Save) Save button save the input values and properites of data structure.

### 4.2.2 Hardware and Sofware Platfrom

The platfrom on which the tool has actually being tested, implemented and tested platfrom is described below.

**Processor :** AMD A6-3420 M APU with Radeon(tm) HD Graphics 1.50 GHz

**RAM :** 4 GB

**OS Platfrom :** Microsoft Windows 7 Home Premium, 64 bit OS

**Software Platfrom :** Eclipse Jee Juno SR1 Win 32

**Language :** Jave 1.6.0; Jave Hot Spot(TM) Client VM 1.6.0-b105

### 4.2.3 User Session

In this section, we will show the typical steps to be followed by the user of the tool to make the circuit, edit, convert, and save the circuit. User have the following steps:

(A). Circuit Entry: The user is ready to form the circuit schematic. Schematic circuit has been divided into time slots. In the user section we assume following circuit is shown in figure 4.2 that is the time slotted version.

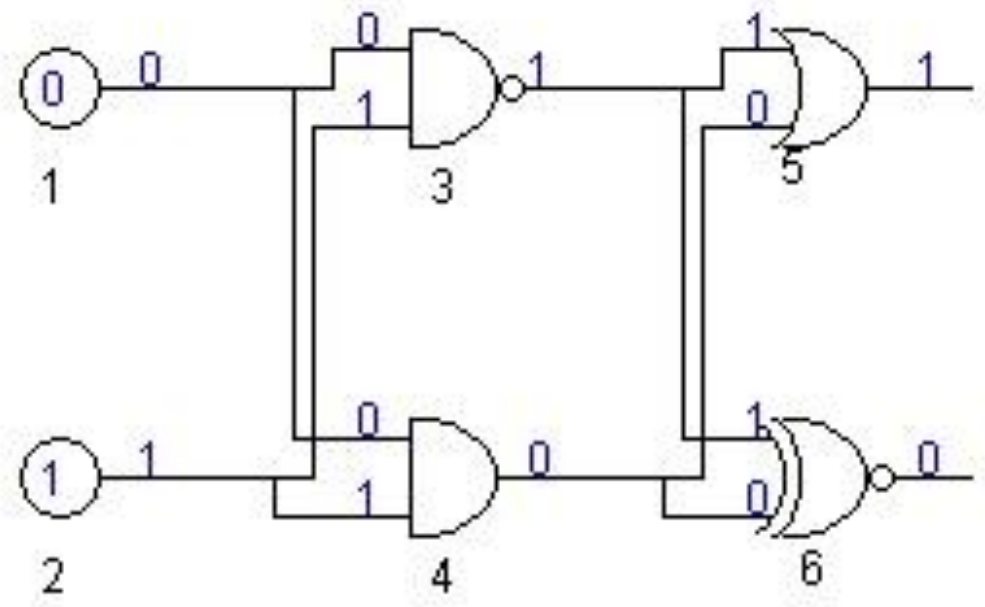

Figure 4.2 : Irreversible Circuit

(B). Entering the Circuit: We accessing the circuit then first we take the input values after selecting the input button, it could be between 0 or 1 than input will display on the top left corner of the display window. All selected input, gates and output first will display on the top left corner block. Immediately we will drag through mouse that input in timeline and slots. Now we can add more gate in the circuit. By selecting each single gate/inputs we can set their properties and also set their reference of every gate. After adding each gate in circuit and set their properties. After that we can click on draw button so we will draw the wiring connectivity of circuit according to the input reference. Circuit diagram is shown in above figure 4.2.

(C). Editing in Circuit: If we want to change gate number 4 . We want to change this gate by replacing another gate than we will simply click on the gate no. 4 (NOR gate), and we have delete (X) button that immediately delete the gate by simple click. After that we will draw the circuit wiring again and we will make the whole circuit connected. Then change the property of that gate and save it. We can see here the edited circuit in figure 4.3.

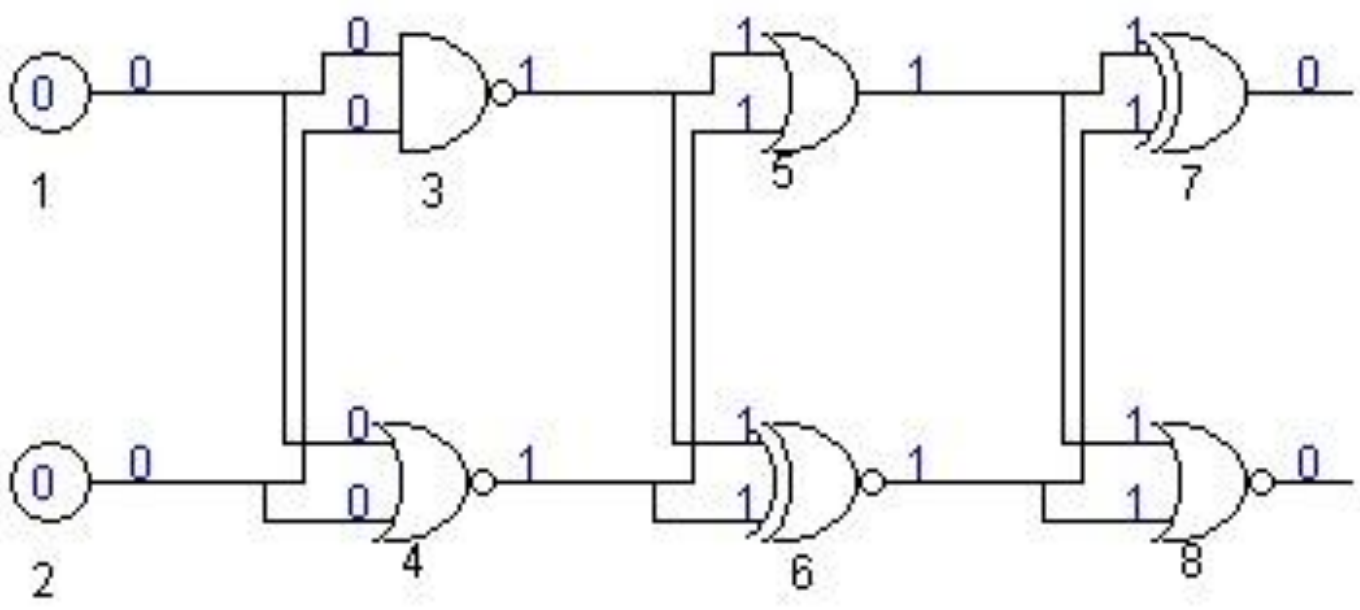

Figure 4.3 : Edited Irreversible Circuit of Figure 4.2

(D). Drawing Circuit: In tool we have draw button placed. This button is used to draw the wiring connectivity between gates. It will connect the reference of gate and draw the connecton of wire as shown in above figure (4.2 and 4.3).

(E). Conversion : In this tool we have convert button placed. This button will convert the irreversible circuit to reversible circuit. This conversion we have every irreversible gate converted into reversible gate. Every gate has their replacement gates in reversible. Converted circuit shown in figure 4.4.

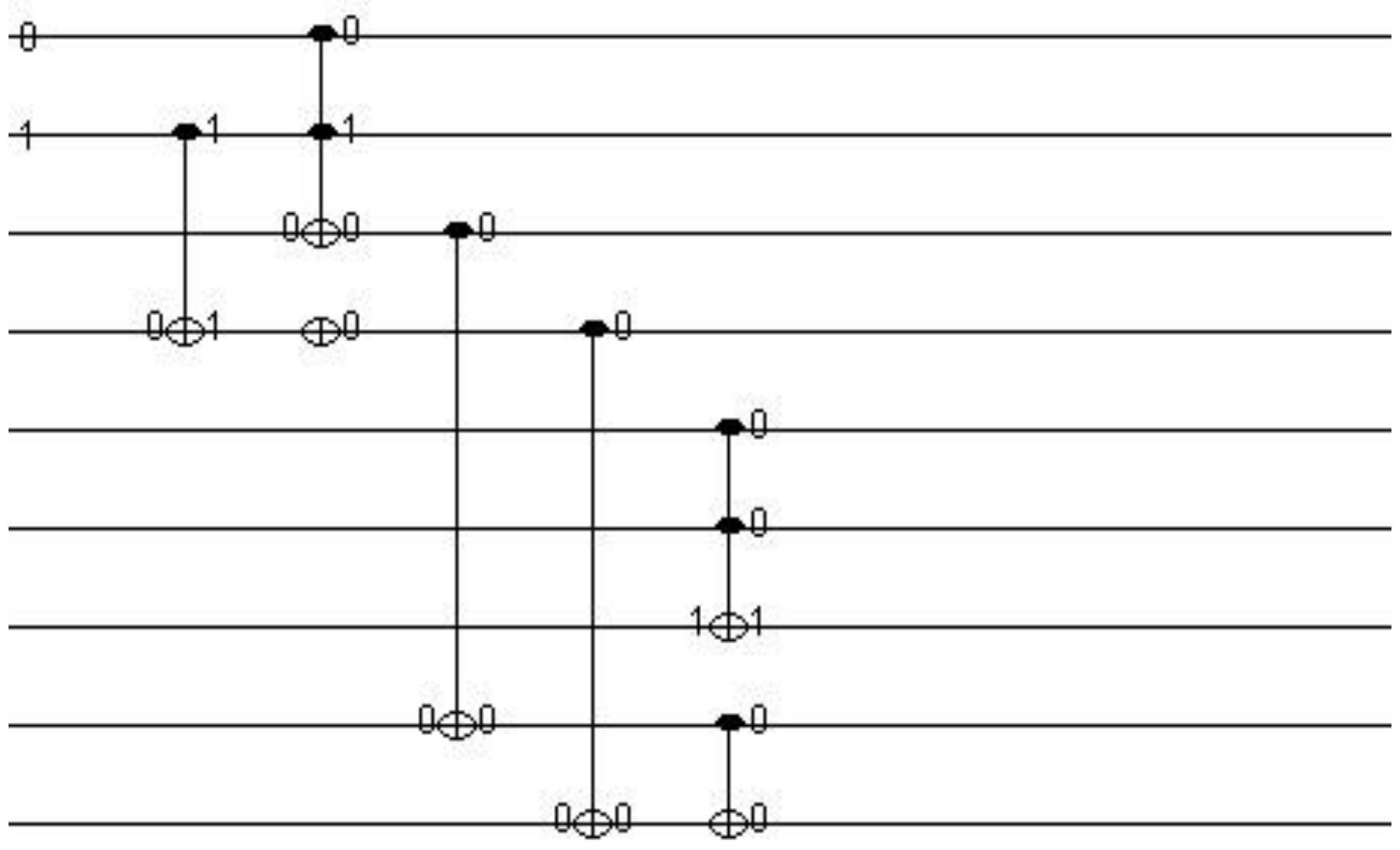

Figure 4.4: Converted Circuit of Irreversible Circuit to Reversible Circuit

## 4.3 Result and Testing

This section included results of some of the circuits for the purpose of verification of the tool.

### 4.3.1 Gate Level Testing

In order to ascertaining the correct working of the tool. First individual gate was verified. Irreversible portion of reversible truth table was verified at individual gate level. This was done adding single gate of circuit in the design and its conversion. The conversion of each implemented gate was individual verified and found correct. The table 4.1 shows the successful verification of the gate level testing.

Table 4.1: Results of Gate Level Testing

| Sr No | Test Case | Truth Table Verified | Test Result |
|---|---|---|---|
| 1. | NOT Gate | Yes | Test Successful |
| 2. | AND Gate | Yes | Test Successful |
| 3. | OR Gate | Yes | Test Successful |
| 4. | NOR Gate | Yes | Test Successful |
| 5. | NAND Gate | Yes | Test Successful |
| 6. | XOR Gate | Yes | Test Successful |
| 7. | XNOR Gate | Yes | Test Successful |

### 4.3.2 Circuit Level Testing

Circuit Level Testing includes the developing of a circuit with their equivalent conversion in reversible. We have already discussed the conventional and reversible circuit in previous chapters. Here we have discussed the converted circuit in reversible and their properties so that we can prove their circuit level testing. In order to faithfully perform design entry, slot wise output is also displayed. A number of circuits have been tested and verified. Two of them are being described below.

- **Half Adder:** The irreversible circuit of half adder is shown in figure 4.5.

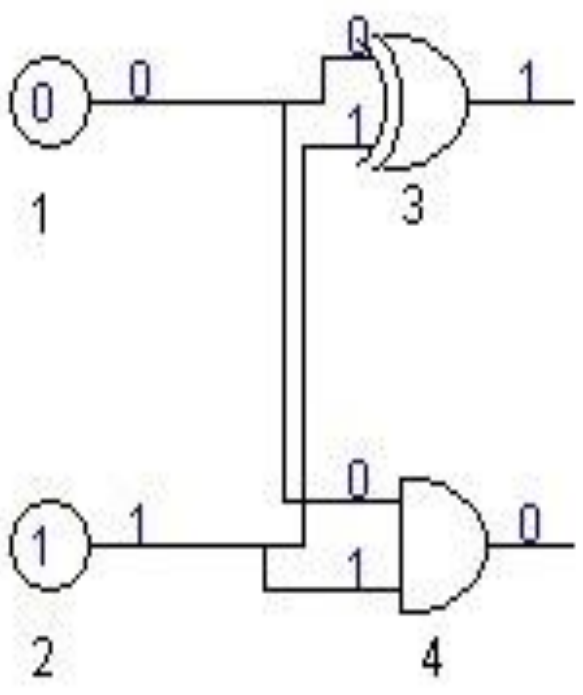

Figure 4.5 Irreversible Half Adder

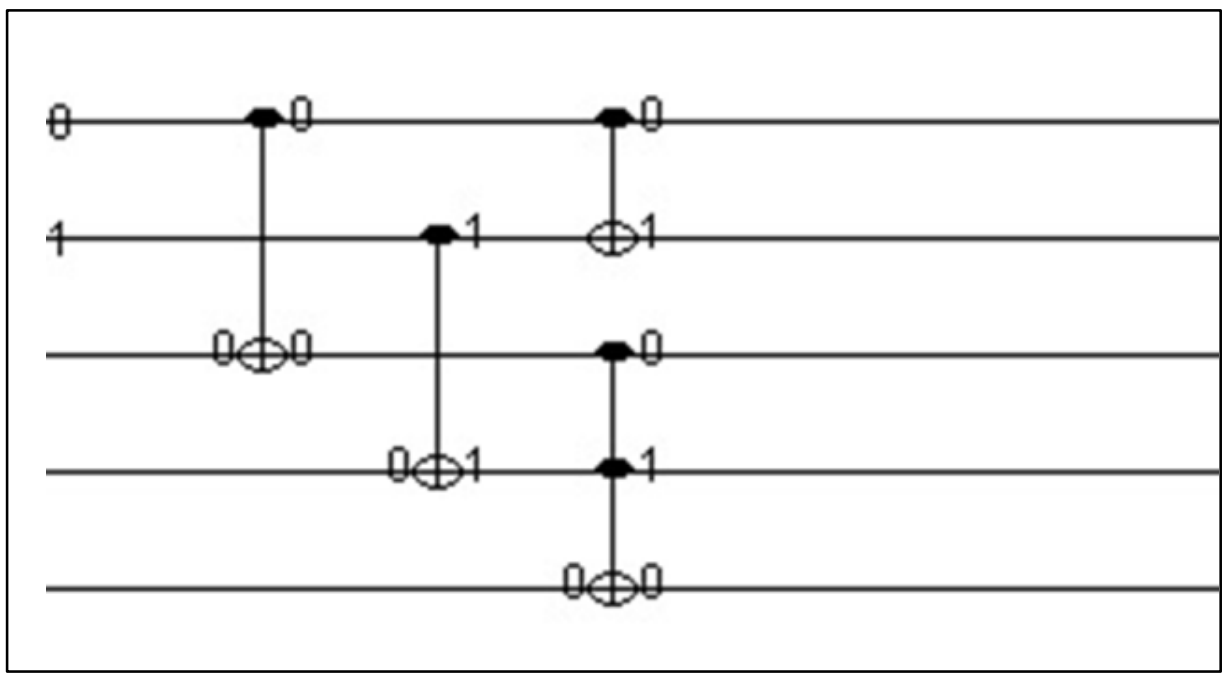

Figure 4.6 : Converted Reversible Circuit of Half Adder

A half adder circuit contains one xor gate and one and gate. In the converted circuit we have done gate by gate conversion. XOR gate can be replaced by CNOT gate and AND gate can be replaced by Toffoli gate. In the above diagram we have shown the gate conversion of both the gate.

- **Arbitrary Circuits:** Here another arbitrary circuit was created and irreversible gates are displayed in figure 4.7.

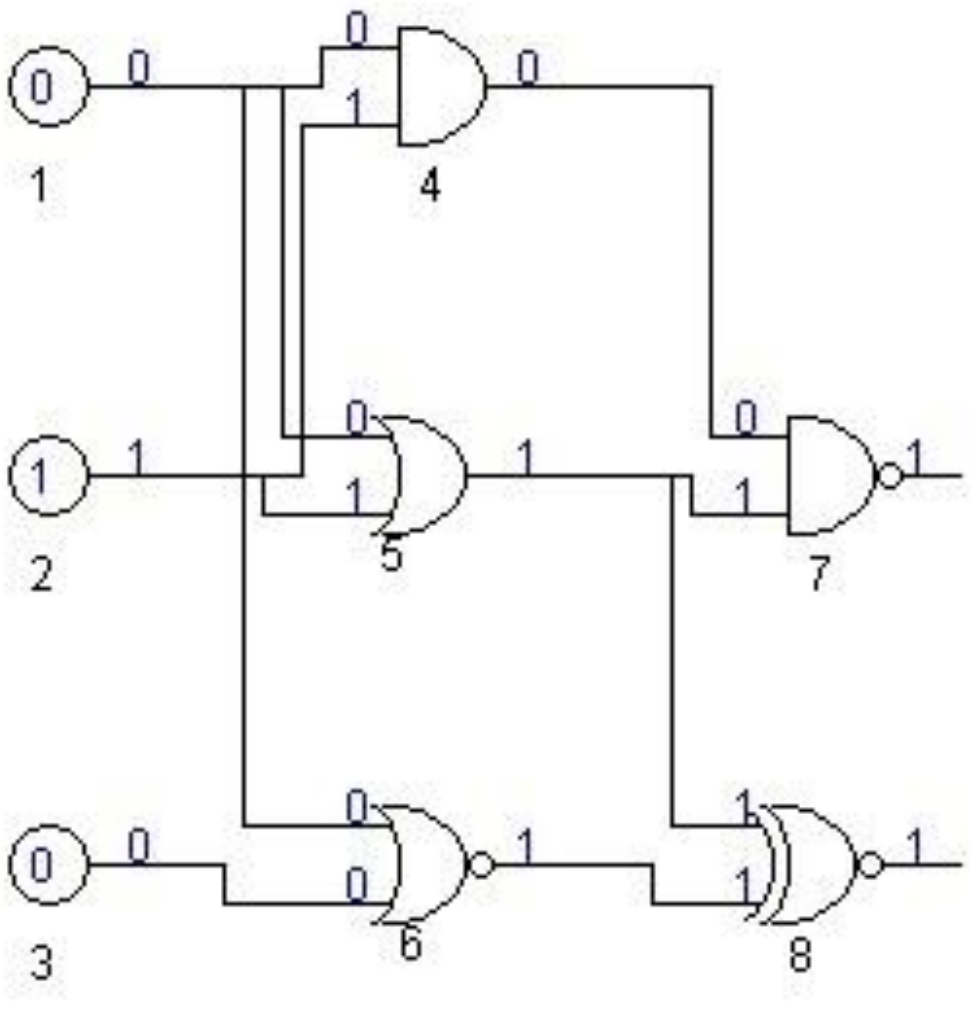

Figure 4.7: Tool Generated Arbitrary Circuit

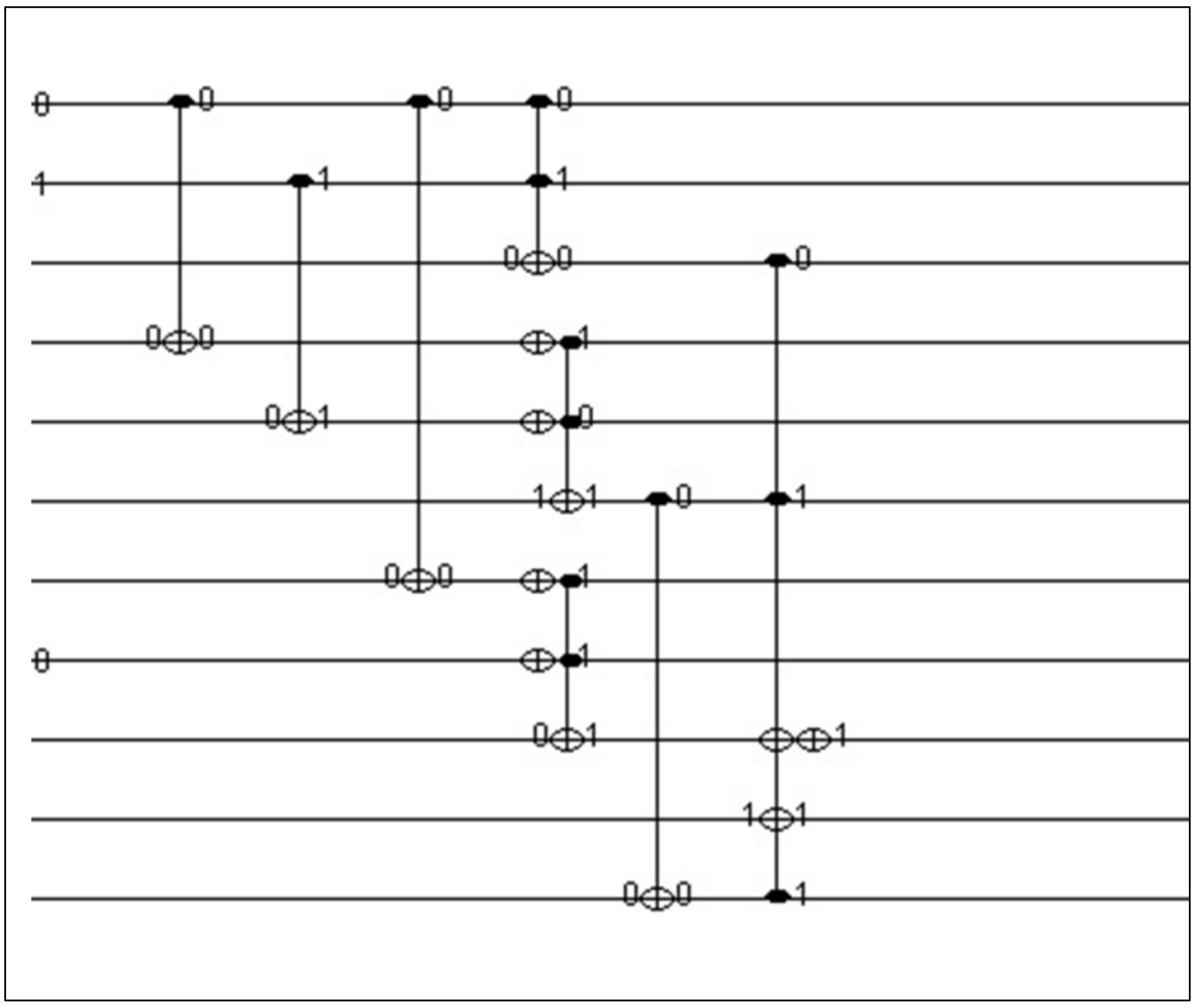

Figure 4.8: Output of Tool Generated of Arbitrary Circuit

As the number of gates are five. The conversion of them was also complex and each gate was replaced by corresponding reversible equivalent gate. A complex reversible circuit is obtained and displayed in figure 4.8. After conversion we observe that functionality of the converted reversible circuit was the same as irreversible circuit.

As verification of above circuits have been found satisfactory. We feel that the algorithms have been faithfully implemented.

# Chapter 5

## CONCLUSION AND DIRECTION OF FUTURE WORK

With emeregence of reversible computing and its promise of low power computation, the usage of old circuit was in focus of this work. A tool named “IRC2RC” for mapping Irreversible circuits to Reversible circuits was developed containing the following specific contributtion.

- Creation of schematic using standard libaries
- Simulate the developed circuit to view output in reversible format
- Editing in Irreversible circuit
- Drawing wiring connvectivity in Irreversible circuit
- Conversion to Irreversible to Reversible circuit
- Save and retrieve developed circuit
- Providing standard circuits as ready example design capture

### Directions of Future Work

The developed tool is now able to capture irreversible design entry and provide reversible logic design. This tool can be further improved by:

- This can also be extended for converting the design in quantum/multivalued circuit.
- Further extension is possible by interfacing with harware description languages.
- It may also be extended to take multivalued and other design formats.
- Further wiring management and optimization can be improved.
- Saving in more than one more format.